\documentclass[twocolumn]{aastex63}
\usepackage{amsmath}
\journalinfo{}

\shorttitle{Planetary Envelope Growth in 3D}
\shortauthors{Bailey et al.}
\graphicspath{{./}{figures/}}

\begin{document}
\title{A Systematic Study of Planetary Envelope Growth with 3D Radiation-Hydrodynamics Simulations}

\correspondingauthor{Avery Bailey}
\email{avery.bailey@unlv.edu}

\author{Avery Bailey}
\affiliation{Department of Physics and Astronomy, University of Nevada, Las Vegas, 4505 South Maryland Parkway, Las Vegas, NV 89154-4002, USA}
\affiliation{Nevada Center for Astrophysics (NCfA), University of Nevada, Las Vegas, NV, USA}

\author{James M. Stone}
\affiliation{School of Natural Sciences, Institute for Advanced Study, Princeton, NJ 08544, USA}
\affiliation{Department of Astrophysical Sciences,
Princeton University, Princeton, NJ 08544, USA}
\author{Jeffrey Fung}
\affiliation{Department of Physics \& Astronomy, 
Clemson University, SC 29634, USA}


\begin{abstract}
In the core accretion model of planet formation, envelope cooling 
regulates the accretion of material and ultimately sets the 
timescale to form a giant planet. Given the diversity of 
planet-forming 
environments, opacity uncertainties, and the advective transport of 
energy by 3-dimensional recycling flows, it is unclear whether 
1D models can adequately describe envelope structure and accretion 
in all regimes. 
Even in 3D models, it is unclear whether approximate 
radiative transfer methods sufficiently model envelope cooling 
particularly at the planetary photosphere. 
To address these uncertainties, we present a suite of 3D radiation 
hydrodynamics simulations employing methods that directly solve 
the transfer equation. We perform a parameter space study, 
formulated in terms of dimensionless parameters, for a 
variety of envelope optical depths and cooling times. We find 
that the thermodynamic structure of the envelope ranges from adiabatic to isothermal based on the cooling time 
and by extension, the background disk temperature and density. 
Our models show 
general agreement with 1D static calculations, suggesting a   
limited role of recycling flows in determining envelope structure.
By adopting a dimensionless framework, these models can 
be applied to a wide range of formation conditions and assumed 
opacities. 
In particular, we dimensionalize them to the case 
of a super-Earth and proto-Jupiter and 
place upper limits on the 3D mass accretion rates prior to runaway growth.
Finally, we 
evaluate the fidelity of approximate radiative transfer methods and 
find that even in the most challenging cases, more approximate 
methods are sufficiently accurate and worth their savings 
in computational cost.
\end{abstract}

\section{Introduction}
Because the growth of planetary envelopes is regulated by cooling, 
an adequate treatment of radiative processes is necessary for 
any comprehensive model of planet formation. In this section, 
we improve our planetary envelope models by including 
radiative transfer. Unlike our previous isothermal models, 
this allows us to study the accretion process and 
quasi-static contraction of the envelope. 1D models of core accretion 
predict various accretion rates that are then used as the
inputs for population synthesis models and compared to 
exoplanet statistics. While these models are good in the sense 
that robustly produce giants at intermediate distances, 
they are not perfect and tend to 
under-produce intermediate mass planets, close-in super-Earths,  
and giants at extended distances. This suggests some problem 
with the underlying assumptions or a currently unaccounted piece 
of physics. Because the inputs are based upon scaling laws and 
solutions obtained from 1D quasi-static models, it is worthwhile to 
ask whether the same results are obtained from 3D hydrodynamics 
models. While this is not the first study to do so 
\citep{AyliffeBate2009,AyliffeBate2012,Schulik+2019,Lambrechts+2019}, 
few studies 
focus on the low mass $\sim 10M_\oplus$ regime. The studies 
that do \citep{DangeloBodenheimer2013,LambrechtsLega2017}, 
recover something similar to 1D models but tend to 
focus on a small number of tailored models and employ 
flux-limited diffusion for the radiative transfer 
(the exception being \citet{Zhu+2021}). Because 
cooling and consequently mass accretion is determined by the 
radiative transfer, flux-limited diffusion may not be appropriate 
in all cases. For these reasons, we find it worthwhile to 
examine the cooling of low mass planets with 
accurate radiative transfer and also a wider parameter space. 
Because we are interested in the formation of a diversity of 
planets at all manner of distances, we will formulate a more 
agnostic dimensionless approach that is more suitable than running 
just a $10 M_\oplus$ Jupiter core at $5$ AU. 

These simulations will also be useful for testing 
various thermodynamic assumptions adopted by 3D hydrodynamics 
studies. In the simplest cases, an isothermal or adiabatic equation 
of state is adopted. Lacking radiative cooling, these simulations 
quickly arrive at a steady state, making for a convenient study of 
the time-independent flow field and associated envelope structure 
\citep{Fung+2015,Ormel+2015,Fung+2019}. Our radiative 
simulations are able to cool and therefore should fall somewhere 
in between the adiabatic and isothermal cases. With our simulations, 
we can measure how isothermal or adiabatic a given envelope is 
and define regions of parameter space where these simple 
thermodynamic assumptions are approximately satisfied. 
Because the adopted thermodynamics are linked to 
kinematic properties of the planet's envelope and things like 
CPD formation \citep{Fung+2019}, this association 
will also help map the kinematic features of planetary 
envelopes into the space of models with realistic cooling.

Our radiative models will also test the validity of cooling 
models that go beyond the simple adiabatic/isothermal assumption.
Some models for example
\citet{Kurokawa+2018}, opt for a simple Newtonian-like cooling 
prescription. This is a computationally expedient 
cooling implementation and useful for constructing toy models of 
planet growth, but it is not self-consistent or physically 
realistic and will fail in regimes to which it is not tuned. 
Other studies adopt flux-limited diffusion for including 
radiative effects in 3D planet simulations 
\citep{DangeloBodenheimer2013,Cimerman+2017,Szulagyi2017,LambrechtsLega2017}.
While FLD is convenient for adding minimal computational cost
and is accurate in the diffusion limit, it 
adopts a heuristic to limit the flux in optically thin regimes. 
Because of this, the method is poorly suited for determining 
radiation fluxes at intermediate $\tau\sim 1$ optical depths.
Being a diffusion approximation, it also assumes that the 
radiative flux is always aligned with the temperature gradient, an 
assumption at odds with the free-streaming of 
radiation in optically thin regions.
These shortcomings have spurred some interest in applying higher 
order moment methods like M1 to the study of planet formation 
\citep{MelonFuksman+2021}. Because we have implemented radiative 
methods into \textsc{athena++} that solve the actual 
discretized transfer equation, our planet models here should test the 
fidelity of these more widely used  
approximate methods. While this is not the primary motivation for 
this study, it should help to inform future studies about the 
relative cost-benefit in moving to more accurate methods of 
radiation transfer. 

The primary goal of this paper is to detail and test the fidelity of 
the method while presenting some properties of the envelope models 
themselves. More physically motivated applications of these models 
are and will be the subject of additional papers \citep{ZhuBailey2023, BaileyZhu2023}.
We begin the rest of this chapter by describing simulation 
details specific to the inclusion of radiative transfer. 
This includes the definition of 
dimensionless parameters used to characterize our radiative models. 
We then thoroughly investigate a single proto-Jupiter model to 
develop an understanding for a prototypical radiative model and 
also act as a test-bed for numerical subtleties. The following 
section expands upon this single model by exploring a wider 
parameter space under different assumed opacities and disk conditions. 

\section{Simulation Details}\label{sec:radsim}
The appropriate equations to 
model 3D envelope evolution are the hydrodynamics equations 
supplemented by source terms due to gravity, rotational motion, and 
radiative cooling. To good approximation, a static form of the 
radiative transfer equations can be used, where terms of order $v/c$ 
or higher are neglected. In a dimensionless form with length scale 
$H_0\equiv \sqrt{kT_0/\mu m_p}$, timescale 
$1/\Omega_0\equiv\sqrt{a^3/GM_\ast}$, and density scale $\rho_0$, 
the equations solved (with coriolis and centrifugal source term 
omitted for notational ease) are:
\begin{align}\label{eq:den}
\frac{\partial \rho}{\partial t} +\nabla\cdot \left(\rho \bm{v}\right) = 0 \\
\frac{\partial \left(\rho \bm{v}\right)}{\partial t} +\nabla\cdot \left(\rho \bm{v}\otimes \bm{v} + p\right) = -\rho\nabla\Phi \\ 
\frac{\partial E}{\partial t} +\nabla\cdot \left(E\bm{v}+ p\bm{v}\right) = -\rho \bm{v}\cdot \nabla\Phi + \rho\kappa\beta
(J - S) \label{eq:energydim} \\
\frac{\partial I}{\partial s} = \rho\kappa \left(S-I\right)
\end{align}
where
\begin{align}
E = \frac{p}{\gamma -1 } + \frac{1}{2}\rho v^2\\
J = \frac{1}{4\pi}\int I d\Omega\\
S =\left(\frac{P}{\rho}\right)^4\label{eq:source}
\end{align}
This particular form 
makes a number of assumptions including an LTE Planck source function,
single (gray) absorption opacity. 
To solve the radiation-hydrodynamics equations, we use the short 
characteristics (SC) method \citep{KunaszAuer1988,AuerPaletou1994} 
identical to that of \citet{Davis+2012}. We also employ a local 
approximation and perform our simulations in a Cartesian coordinate 
system centered on the planet. Code units 
are taken by setting $H_0=\Omega_0=\rho_0 = 1$, reflecting the above 
choice of dimensionless equations.
With this setup, all models have the same initial equilibrium state:
$\rho=p=\exp(-z^2/2)$, and a y-velocity shear in $\hat{x}$. 
This equilibrium holds for both an adiabatic and isothermal 
equation of state, so long as no planet is introduced. 
With 
radiative transfer, the incident intensity at each boundary 
is fixed to the value of the initial source function (unity in code 
units) so that with no planet, the disk exists in radiative 
equilibrium. Physically this incident intensity can represent 
irradiation by the star or upper layers of the protoplanetary disk. 
In the future, the multi-frequency capabilities of the code could 
be leveraged to treat the incident radiation as ultraviolet but for 
now we limit ourselves to the single frequency irradiated case.
With an initial equilibrium state, we 
introduce the planetary potential and investigate the 
resultant evolution under the influence of radiative cooling. 
This is contrived in the sense that real planets are not ``suddenly 
introduced'', they evolve in conjunction with the disk throughout the
formation process. Nevertheless, this methodology is consistent with 
other radiative-hydro simulations
\citep{AyliffeBate2012,DangeloBodenheimer2013,Cimerman+2017,LambrechtsLega2017}. 

To make the potential parameter space more tenable, we adopt fixed 
vales of $\mu = 2.3$ and $\gamma = 1.4$ between our models.
With $\mu$ and $\gamma$ fixed, 4 dimensionless 
parameters remain to uniquely characterize a given model. 
Two of these, thermal mass $q_t$ and softening length 
$\epsilon$ are contained within the adopted form for gravity:
\begin{equation}
\Phi = \frac{GM_p}{\sqrt{r^2+\epsilon^2}}\left(\frac{1}{\Omega_0^2 H_0^2}\right) = \frac{q_t}{\sqrt{(r/H_0)^2 + (\epsilon/H_0)^2}{}}
\end{equation}
Whereas $q_t$ is a parameter with physical significance, $\epsilon$ 
is less physical, arising partially out of numerical necessity.
For another dimensionless parameter, we have defined $\beta$. Defined as 
\begin{equation}
\beta \equiv \frac{ac\left(\frac{\mu m_p}{k}\right)\frac{T_0^3}{\rho_0}}{c_s}\sim \frac{c_\gamma}{c_s}
\end{equation}
is roughly the
ratio of the characteristic velocity of photon diffusion $c_\gamma$ 
relative to the disk sound speed $c_s$. A fourth dimensionless 
parameter enters through the choice of opacity law. Our models will 
assume a constant opacity $\kappa_0$, so that the dimensionless 
opacity is simply 
\begin{equation}
\kappa = \kappa_0 \rho_0 H_0 \approx \kappa_0 \Sigma_0,
\end{equation}
and is therefore roughly the disk's vertical optical depth. 
With these four dimensionless parameters, $q_t$, $\epsilon$, $\beta$, 
$\kappa$, a given model can be uniquely expressed. This use 
of dimensionless parameters will simplify the discussion of 
planet formation when it comes to comparing models representing 
different planet-forming environments. 
We note that this whole 
methodology and creation of 4 dimensionless parameters neglects 
several potential pieces of microphysics that would 
otherwise add to the complexity of the models here. For example, 
we neglect ionization and molecular dissociation of Hydrogen 
which can change the equation of state and serve as an 
energy sink. While these are important, they are higher order 
corrections to our models especially in the context of our overly 
simple constant opacity assumption.

\section{A Proto-Jupiter Model ($\beta=1$, $\kappa=100$)}
With the model details defined, we focus on a prototypical 
simulation of a Jupiter-like core under reasonable conditions at 
$5$ AU in a protoplanetary disk. For this model we adopt the
dimensionless parameters $q_t = 0.5$, $\beta=100$, $\kappa=1$ which 
corresponds to roughly a $10M_\oplus$ core orbiting at $5$ AU around 
a solar mass star in a disk with background temperature $T_0=70$ K, 
integrated surface density $\Sigma_0=250$ g/cm$^2$, and opacity
$\kappa_0=1$ cm$^2$/g. With this opacity, the entirety of the 
Bondi sphere is optically thick, with the photosphere occurring near 
the vertical boundary of our simulation box at $2H_0$. 
This choice of parameters also corresponds to scale height aspect 
ratio of $h_p=0.04$ at 5 AU. 
For now, we set the gravitational 
softening length at $\epsilon=0.1 R_b$, comparable to similar 
studies \citep{LambrechtsLega2017,Lambrechts+2019,Schulik+2019}.

Though we have championed the short characteristics implementation 
of radiative transfer in Section \ref{sec:radsim}, 
the transition to a regime in which 
this method starts to become computationally inefficient coincides 
with proto-Jupiter conditions. For this reason, our fiducial model in 
this section is performed with the method of \citet{Jiang2021}. 
We note that the method of \citet{Jiang2021} is not solving the same 
equations as those in Section \ref{sec:radsim} as it retains 
velocity dependent terms in the Lorentz transformation for the 
radiation. In the low-velocity limit, of course, 
this method does reduce to our static limit. For notational purpose 
we will sometimes refer to the \citet{Jiang2021} as the `implicit' 
method and SC as `explicit' referring to the manner in which the 
transfer equation is solved.

For this proto-Jupiter case,
we do run an additional SC model (albeit for fewer orbits) 
and confirm that 
the two radiation schemes return nearly indistinguishable results. 
Our fiducial model is run for 
$15$ orbits and adopts a generous angular discretization of $12$ 
angles per octant. We use two levels of mesh refinement so that the 
highest resolution attained is 128 cells/$R_b$ and the softening 
length is resolved by $12$ cells. 
With these fiducial parameters, 
a simulation takes roughly 6000 core-hours to run 15 orbits. While 
mesh refinement makes it easy to increase the resolution without 
needing to add many more processors, more refinement levels also 
leads to more iterations of the radiation solver. This is because 
each processor solves its own local radiation problem on a part of 
the mesh and then communicates and iterates with other processors 
until a good enough global solution is found. Also, because the 
gravitational acceleration scales rather steeply with the softening 
length, it tends to be hard to get converged results for small 
softening length. For these reasons, in this work we focus on 
studying a number of models we trust are converged rather than 
1-2 expensive high resolution models subject to more uncertainty. 

The simulation begins from equilibrium with the introduction of 
the planetary potential. This triggers a rapid infall of gas as the 
atmosphere attempts to establish a pressure gradient in response 
and arrive at a new equilibrium. 
Within a single orbit, a spherically symmetric envelope forms around 
the planet and an approximate hydrostatic equilibrium is attained. 
Gas that has fallen into the planetary potential is heated, giving 
the envelope a temperature gradient. Due to the influence of 
radiative cooling, the temperature gradient is not as severe as an 
adiabatic profile. The gas is heated to only several times the 
background disk temperature because of our large softening length 
and absence of heating sources. In analytic models, this outer part 
of the planetary envelope is typically regarded as isothermal 
\citep{Rafikov2006,PisoYoudin2014}, reflecting the modest 
temperature increase in this model. 

\begin{figure*}
\begin{center}
\plotone{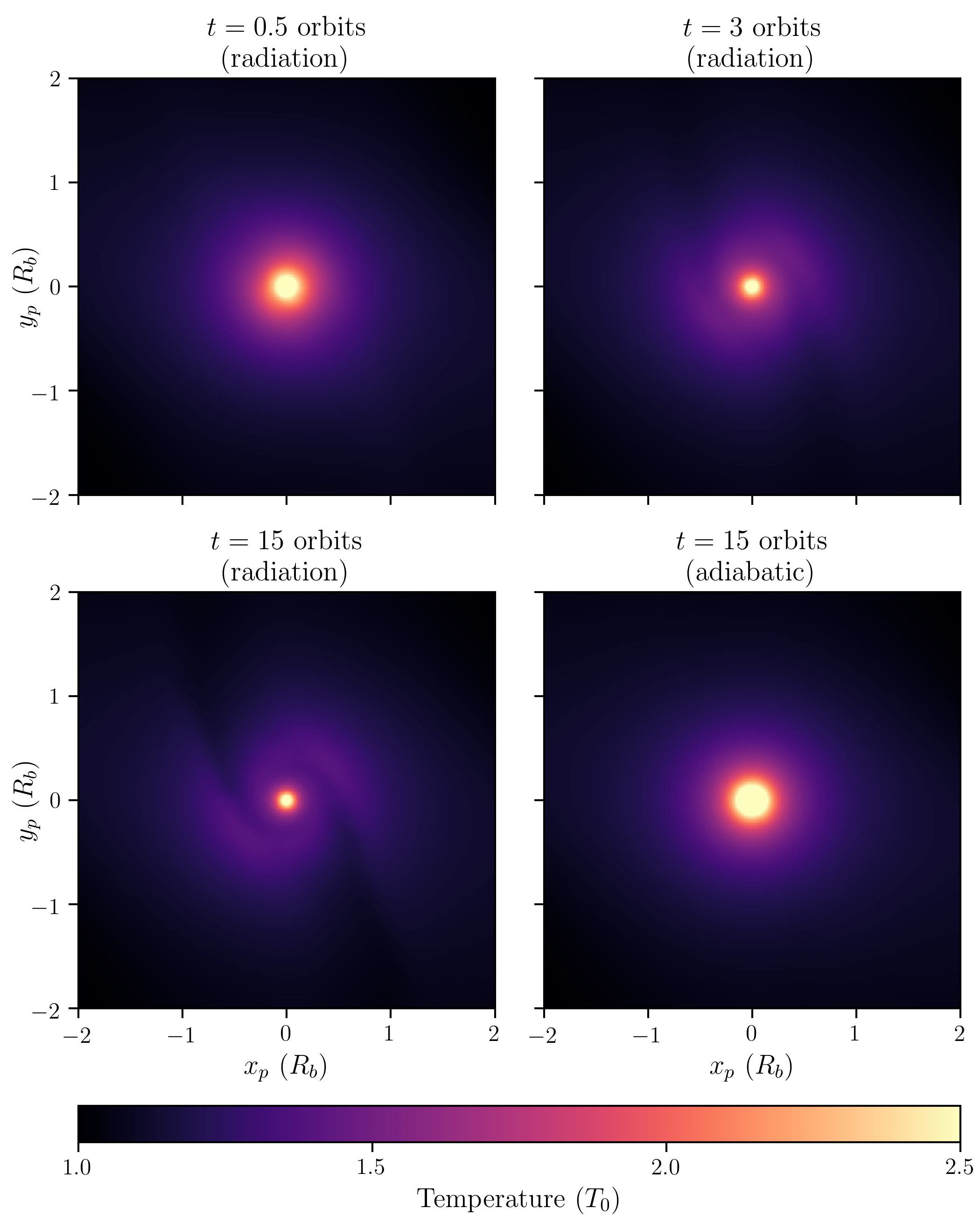}
\caption{Evolution of the midplane temperature in the fiducial 
proto-Jupiter simulation. The upper left panel shows the spherically 
symmetric profile at $0.5$ orbits soon after the planet has been 
introduced. The upper right is a snapshot at $3$ orbits. At this 
point, the envelope has largely adjusted to the planetary potential 
and radiative cooling is governing the subsequent evolution. The 
lower left panel shows the further cooled profile after $15$ orbits.
The lower right panel shows the adiabatic steady state for comparison.
Cooled substructures and departures from spherical symmetry are 
more prominent in radiative models.}\label{fig:juptemp}
\end{center}
\end{figure*}

After reaching a state of near hydrostatic equilibrium, the envelope 
cools radiatively and the temperature profile falls monotonically 
with time. 
This cooling leads to departures from spherical 
symmetry in the envelope structure. Near the midplane, cool spiral
substructures appear, aligned with the boundary between outgoing 
horseshoe flows and background shear flows. These 
departures from spherical symmetry can be seen in Fig.\@
\ref{fig:juptemp} where we plot snapshots of the midplane 
temperature distribution at different times in the fiducial 
simulation. While spiral arms are generic for 
planets irrespective of their thermodynamics, these cooled 
substructures only appear in radiative models. Furthermore, these 
cool spiral arms are Kelvin-Helmholtz unstable. In our fiducial 
simulations, Kelvin-Helmholtz rolls are continually observed for 
the full 15 orbit duration but can be difficult to discern in 
the still images of Fig.\@ \ref{fig:juptemp}. With higher resolution, 
the instability becomes more vigorous and discernable. 
The continual production of Kelvin-Helmholtz rolls leads to mixing 
and small-scale turbulent velocities in the vicinity of the planet.

\begin{figure}
\begin{center}
\plotone{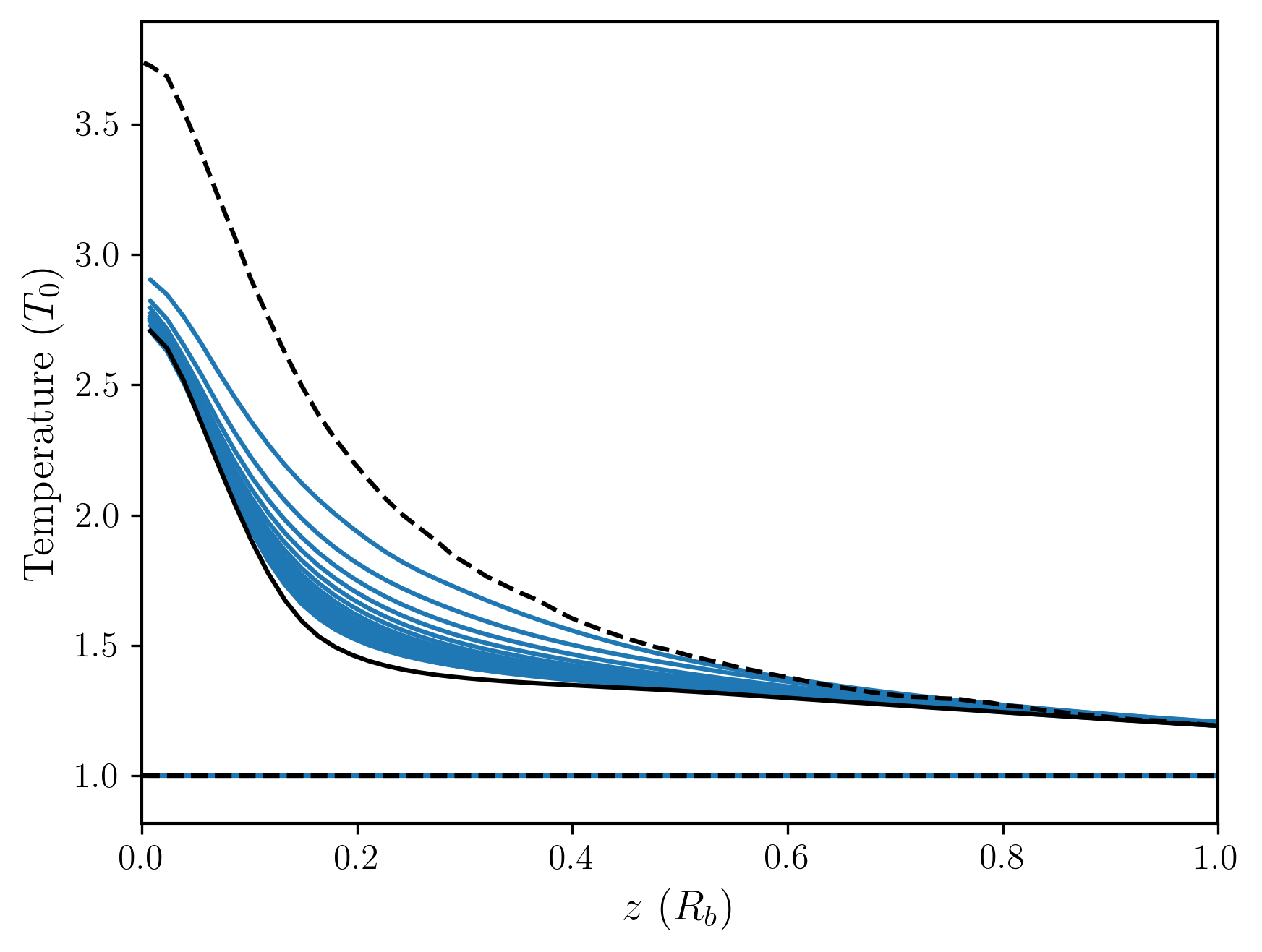}
\caption{Evolution of the vertical temperature profile for the 
proto-Jupiter model. Profiles from the fiducial model are sampled  
along the pole each orbit for 15 orbits and plotted in blue. After 
an initial heating phase during the introduction of the potential 
the profiles monotonically cool. We plot the profile from a longer 
test run at $40$ orbits as the solid black line for comparison. 
Dashed lines show the bounding temperature profiles obtained from 
corresponding adiabatic and isothermal models. }\label{fig:jupvert}
\end{center}
\end{figure}

Envelope cooling slows as the simulation progresses, such that 
by the time the simulation ends the cooling rate has dropped 
by more than an order of magnitude in the deep envelope interior. 
This fact is reflected in the convergence of profiles in Fig.\@ 
\ref{fig:jupvert} where the polar temperature profile is plotted 
every orbit for the fiducial model. Even an exploratory model 
more than doubling the runtime to $40$ orbits shows nearly the 
same temperature profile 
(solid black line in Fig.\@ \ref{fig:jupvert}). 

\begin{figure}
\begin{center}
\plotone{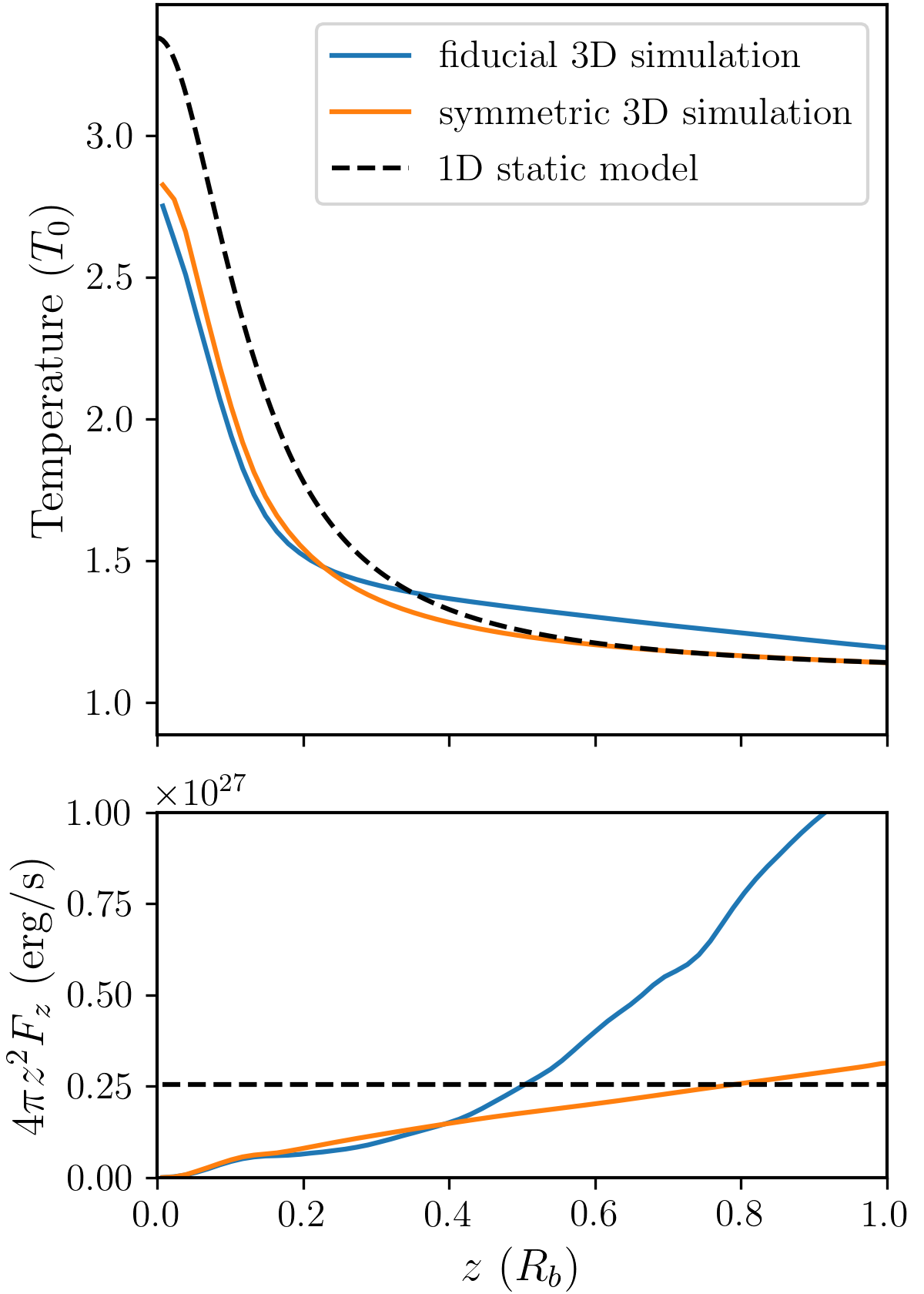}
\caption{Vertical temperature (top) and luminosity (bottom) of 
the envelope in our proto-Jupiter model after 15 
orbits. The blue curve shows our fiducial simulation. The orange 
curve is a 3D rad-hydro simulation but treating only the planetary 
potential so that the simulation is spherically symmetric. 
The dashed curve is a 1D hydrostatic model assuming a constant 
luminosity. The luminosity in all cases is calculated using the 
polar radiative flux $F_z$ assuming spherical 
symmetry.}\label{fig:jup1d}
\end{center}
\end{figure}
Notably, this 
steady cooled profile is distinct from both adiabatic and 
isothermal cases, though it does remain bounded by the two. 
Nevertheless, the temperature profile appears similar to what 
would be obtained from a 1D static model. In Fig.\@ 
\ref{fig:jup1d} we plot our final simulation temperature profile 
against one obtained from a symmetric 3D simulation and a 1D 
static model and find general agreement between the three. 
In the case of the symmetric 3D simulation, we ran the same model 
as the fiducial case but starting from a uniform static medium 
and removing the stratification/rotational source terms. 
Though this model is not truly symmetric because 
the box is still Cartesian with various boundaries,  
inspection indicates that the simulation evolves close 
to spherical symmetry. The 1D static model is identical to those 
of \citet{PisoYoudin2014} but using our static softened potential 
to be more comparable to the 3D simulations. For the boundary 
conditions on the static model we take the temperature and 
pressure at $R_b$ from our symmetric 3D model rather than the 
unperturbed disk values as our models demonstrate that the planet 
does still effect disk at this range. While the 1D profile 
does show minor deviations from the 3D profiles, like a higher 
central temperature, we believe this is primarily due to 
the constant luminosity assumption in our 1D model. As in 
\citet{PisoYoudin2014}, we construct our 1D model by making the 
luminosity constant and tuning it to the exact value as to place 
the envelope in hydrostatic balance. We plot the luminosities of 
each model in the lower panel of Fig.\@ \ref{fig:jup1d} and find 
that temperature profiles diverge as the constant luminosity 
assumption overestimates the luminosity emerging from the 
envelope interior. At the same time, the symmetric 1D and 3D models 
tend have a much lower luminosity in the envelope outskirts. 
While this difference manifests weakly in the temperature profile, 
the higher luminosity indicates that these growing envelopes 
could be more observable that 1D models predict. 

Compared with the temperature profile, the density profile 
shows a similar pattern of relaxation: the density increases 
as the atmosphere accretes, all the while maintaining a 
quasi-static structure. Density profiles also remain bounded 
by the isothermal and adiabatic profiles. As pointed out in 
\citet{Fung+2019}, the envelope masses in adiabatic and isothermal 
models tend to be lower than necessary for runaway accretion. 
It seems that this problem persists in radiative models, at least 
for large softening length. Because models with smaller softening 
length tend to show greater rotation and rotationally supported 
envelopes are allowed arbitrary density profiles, it is possible 
that a rotationally supported radiative model could accrete beyond 
current model limits. While we have run preliminary models with 
smaller softening length to address this question, the increased 
gravitational acceleration seems too large to be compatible with
our radiation solver at the current resolution.

\begin{figure}
\begin{center}
\plotone{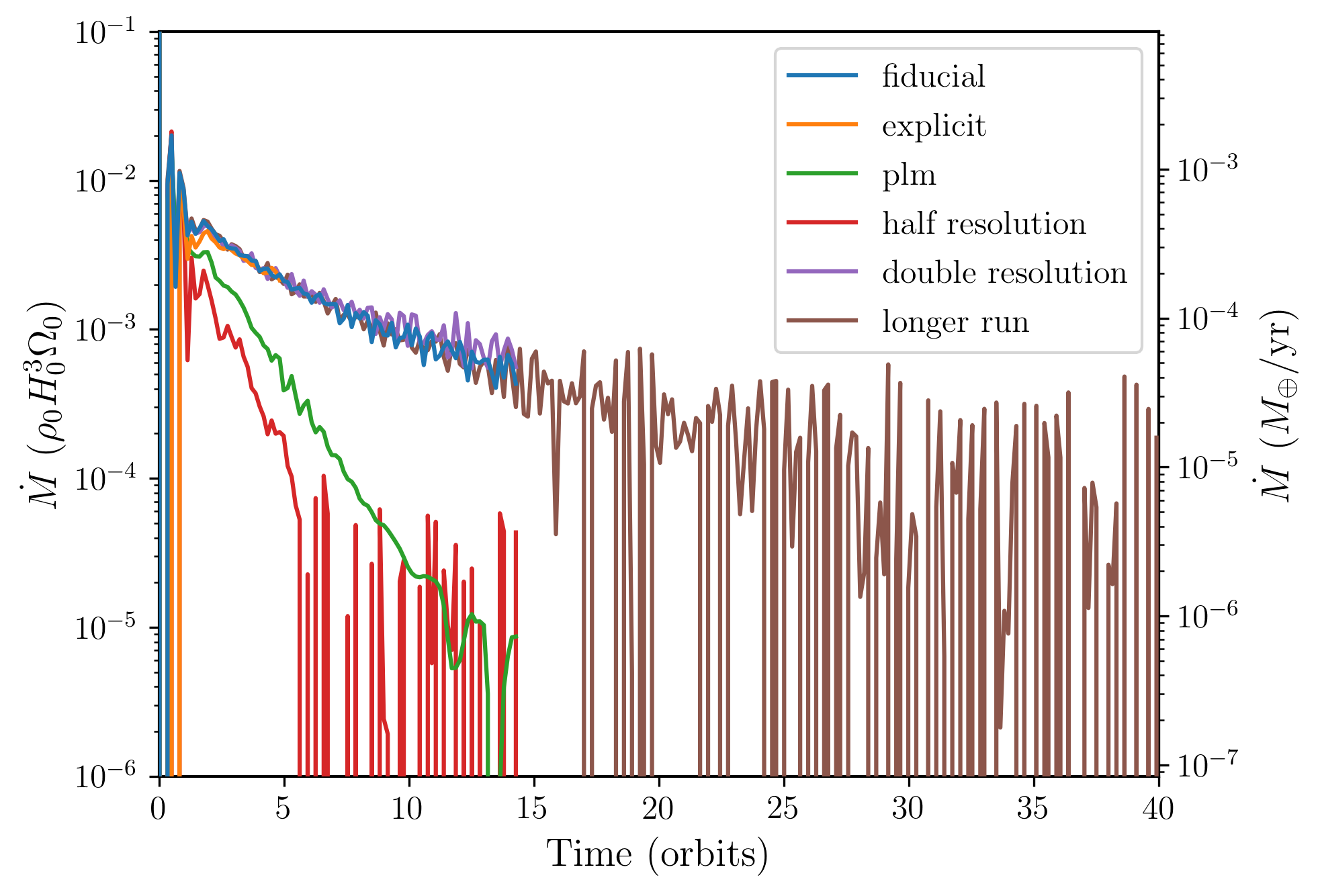}
\caption{Mass accretion rate within the Bondi sphere as a function of 
time for models with parameters representative of a proto-Jupiter. 
The agreement between the fiducial and doubled resolution runs 
demonstrates the convergence of our fiducial model. Poorly resolved simulations tend to underestimate accretion rates as is evidenced by 
the half-resolution and piecewise linear (PLM) reconstruction runs. }\label{fig:jup}
\end{center}
\end{figure}

As cooling of the envelope interior slows, so too does the 
accretion of gas. We show the rate of change of mass within the 
Bondi radius as a function of time in Fig.\@ \ref{fig:jup}. While 
some of the mass contained within the Bondi radius may be unbound 
and recycled with the background disk, 1D models treat this as 
formal size of the envelope, making it a useful as a diagnostic and 
point of comparison. For the timebeing, we treat this change in 
mass as an envelope accretion rate as other authors have done but 
recognize the uncertainty in doing so. By the end of the simulation, 
this accretion rate has dropped to a rate of $4\times10^{-5}$ 
$M_\oplus$/yr while the mass inside $R_b$ has leveled out 
to a value $\approx 0.3 M_\oplus$. 
This appears roughly consistent with similar 3D works 
like \citet{Lambrechts+2019} who extrapolate an accretion rate 
$\approx 10^{-5}$ $M_\oplus/$yr or \citet{AyliffeBate2009} 
who find $2\times 10^{-4}$ $M_\oplus/$yr for a $10 M_\oplus$ core. 

We validate the fiducial model by running a 
host of other simulations with different numerical choices. 
These include: a simulation with the explicit SC radiation method for 
fewer orbits, a simulation with piecewise linear reconstruction, 
a simulation with half-resolution everywhere on the grid, 
a simulation with double-resolution everywhere on the grid, and a longer
simulation run for $40$ orbits but compensating with a fewer angles 
per octant. The accretion rates of these test simulations are also 
plotted in Fig.\@ \ref{fig:jup}. The good agreement between the
fiducial model and the explicit/double-resolution runs suggests that 
the fiducial model is converged. The PLM and half-resolution runs 
demonstrate the finding that under-resolved models to underestimate 
accretion rates. This is consistent with the findings of 
\citet{Schulik+2019}, requiring $\sim 10$ cells per softening length 
to obtain a converged accretion rate. The accretion rate for 
the longer run continues to fall but also becomes significantly 
noisier. The good agreement with the fiducial model at $<15$ orbits
suggests that at least in this optically thick case, 12 angles per 
octant more than sufficient for convergence in the optically thick 
regime.

\section{Diversity of Planet-Forming Environments}
While the four parameters $q_t$, $\epsilon$, $\kappa$, $\beta$ are less 
intuitive than a parameter like the planet mass they are much more 
convenient for uniquely defining models. For example, a $10 M_\oplus$ 
Jupiter-like core ($T_0\approx 70$ K, $\Sigma_0\approx 250$ g/cm$^2$,  
$a\approx 5$ AU, $\kappa\approx 1$ cm$^2$/g), could be described by 
the same model as a $50 M_\oplus$ planet at $a\approx 100$ AU 
(with $T_0\approx 10$ K, $\Sigma_0\approx 100$ g/cm$^2$, $\kappa\approx 
2.5$ cm$^2$/g) because they share the same $q_t$, $\beta$, $\kappa$.
Unless of course some other physics is added, 
e.g. dissociation of molecules, which sets a temperature scale and
breaks this dimensionless scaling.

\begin{figure}
\begin{center}
\plotone{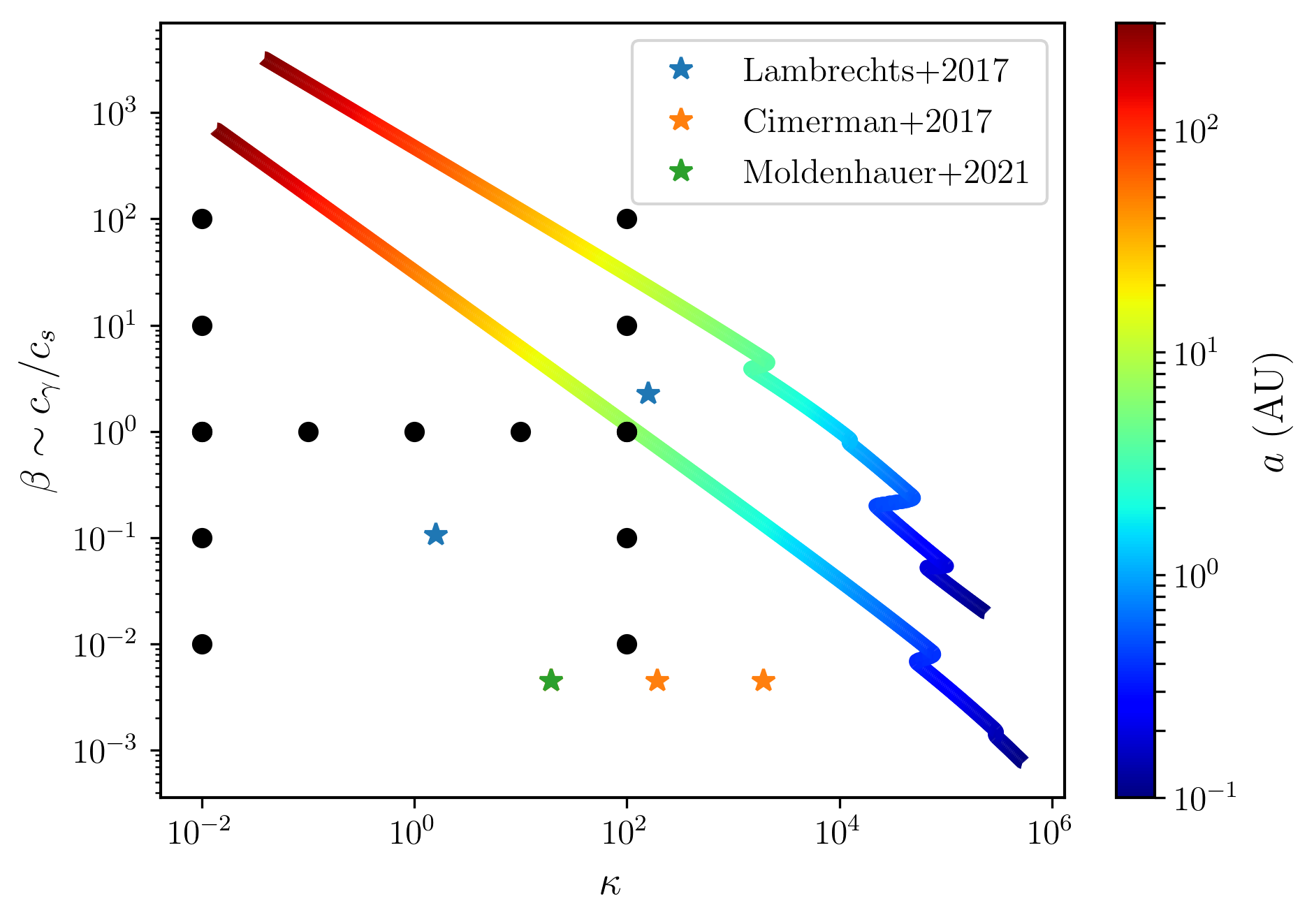}
\caption{Parameter space of $\beta$ and $\kappa$. Our models are 
shown as discrete black dots. Colored lines show the trend of 
parameters as a function radial location in disk for two choices of 
radial temperature profile in the protoplanetary disk. The parameters 
of other studies that investigate planets with similar thermal mass 
are also marked and labelled.}\label{fig:par}
\end{center}
\end{figure}

While the notion of dimensionless parameters simplifies the modelling 
of forming planets, there must be some connection with the physical 
dimensional parameters in order to develop an interpretable  
understanding of planet formation. Here we develop some intuition 
for how reasonably realistic planets translate to our dimensionless 
models and vice-versa. This will be useful for the following section 
where we examine a grid of models spanning a range of 
different dimensionless parameters.
To begin, we can associate dimensionless parameters  
with disk properties like $T_0$, $\Sigma_0$ by assuming 
some reasonable profiles for a protoplanetary disk. For a minimum-mass 
solar nebula (MMSN) structure, 
($\Sigma\propto a^{-3/2}$, $T_0\propto a^{-1/2}$) the dimensionless parameters then scale with just the physical planet parameters 
$M_p$, $a$ and disk opacity $\kappa_0$ as:
\begin{align}
q_t &\propto M_p a^{-3/4}\\
\kappa& \propto \kappa_0(a)a^{-3/2}\\
\beta&\propto a^{3/2}
\end{align}
With some assumed opacity law $\kappa_0(\rho,T)$, the dimensionless 
parameters can be expressed as solely functions of the readily interpretable parameters $a$, $M_p$. This gives  
some intuition about where in the disk a given model should be 
applicable. In both 3D and 1D models of planet formation it is 
typical to adopt the dust grain opacity law of \citet{BellLin1994}:
\begin{equation}
    \kappa_0 = 2\left(\frac{T_0}{100 \text{ K}}\right)^2 \text{ cm}^2\text{/g}\label{eq:belllin}
\end{equation}
While widely used for being analytically tractable, this opacity law 
assumes an interstellar dust-to-gas ratio of 0.01 and 
is most applicable in cold regions of the disk $T< 100$ K. This gives 
a scaling for the dimensionless opacity:
\begin{equation}
    \kappa \propto a^{-5/2}
\end{equation}
Notably, $\kappa$, $\beta$ are functions of the disk properties 
while $q_t$ is a function of both the planet mass and the assumed 
background disk structure. With this we can map the variation of the
two dimensionless parameters $\beta$, $\kappa$ for a given disk model 
as in Fig.\@ \ref{fig:par}. 
In Fig.\@ \ref{fig:par}, we present 
two curves from reasonable protoplanetary disk models in the 
($\kappa,\beta$) plane. Each point on the curves is colored 
according to its radial location in disk. 
The two curves adopt different temperature profiles, with the lower 
one coming from \citet{ChiangYoudin2010} and the upper one from 
\citet{Rafikov2006}. The scalings are nearly equivalent but the 
upper one is a factor of $2.5$ hotter. They both adopt the 
same MMSN density profile. 
In the Figure, we drop the convenient analytic opacity of 
Eq.\@ \eqref{eq:belllin} for \citet{Semenov+2003} opacities 
which include components unique to hotter parts of the disk. Jumps 
in the opacity law cause the sharp features in hotter opaque regions.
The \citet{BellLin1994} law is still reflected in the 
$\beta\propto\kappa^{-3/5}$ scaling at large $a>10$ AU where the disk 
becomes cold. From this plot it is apparent that for a given disk 
profile, moving outwards in the disk corresponds to larger $\beta$ 
and lower $\kappa$. Though the precise normalization and slope 
of this curve can change significantly based upon assumption about 
opacity, disk mass, etc., the general fact that larger
orbital radius corresponds to larger $\beta$ and lower $\kappa$ is 
robust. We also highlight the fact that the spanned parameter 
space throughout the full disk is rather large, going from 
optically thick regimes in the inner disk to optically thin 
in outer disk. 

In general, studies employing 3D radiative models tend to focus on a 
single region of the disk. Usually parameters approximating a 
prototypical Jupiter or super-Earth are selected and then studied 
under some variation of either opacity or planet mass. 
While this is sensible, especially considering the computational 
cost involved, it makes it considerably harder to compare between 
models and isolate the role of the local disk conditions 
$T_0$, $\Sigma_0$, $\kappa_0$ etc.\@ on the formation process. 
To this end, in the next section we perform a series of 
simulations spanning a range of $\kappa$ and $\beta$ and study the 
resulting envelope structure and accretion.

\section{A Suite of Radiative Models}
Figure \ref{fig:par} and the preceding estimates suggest that 
depending on the location in disk, planets could form under a range 
of conditions from optically thin $(\kappa \ll 1)$ to optically thick 
$(\kappa \gg 1)$ but also under a range of $\beta$ spanning above 
and below unity. With this in mind, we attempt to cover this 
($\beta$,$\kappa$) parameter space with a set of models that is 
informative but also computationally feasible. The chosen models 
are marked in Fig.\@ \ref{fig:par} with circles and correspond to 
three different branches --- an optically thin $(\kappa=10^{-2})$ 
branch, an optically thick branch $(\kappa=10^2)$, and a $\beta=1$
branch. Though this configuration results in models that aren't 
necessarily realistic (e.g. $\beta=\kappa=10^{-2}$), it is 
convenient from a theoretical perspective, being symmetric about 
unity and allowing us to test the independent variation of $\kappa$ 
and $\beta$. At the same time, it does cover realistic regions of 
parameter space, in many cases overlapping with the work of 
other studies. By design, our previously presented proto-Jupiter 
model $(\beta=1$, $\kappa=100)$ lies directly on this grid, overlapping 
with the radiative Jupiter-like models by 
\citet{LambrechtsLega2017,Lambrechts+2019,Schulik+2019}. Similarly, 
our $(\beta=10^{-2}$, $\left.\kappa=10^2\right)$ model is roughly 
consistent with the super-Earth conditions adopted by the radiative 
simulations of \citet{Cimerman+2017,Molenhauer2021}. However, these 
``super-Earth conditions'' assume opacities lower by several orders 
of magnitude than the \citet{Semenov+2003} opacities. This is usually 
done to hasten cooling and make the problem more computationally 
tractable while still qualitatively preserving the optically thick 
formation conditions. For this suite of models we fix 
the thermal mass to $q_t=0.5$, a convenient choice applicable to both 
Jupiter-like cores and super-Earths. 
The models presented are run for $15$ orbits. We also fix the softening 
length to $\epsilon = 0.1R_b$, consistent with our aforementioned 
proto-Jupiter model. While this is orders of magnitude too large to 
represent the core of a Jupiter-like planet, the choice is convenient 
for the super-Earth case where it would be comparable to the core 
radius. 

\subsection{Optically Thin Models}
\subsubsection{Envelope Structure}
These models, occupying the left vertical branch in Figure 
\ref{fig:par}, are such low opacity that the optical depth through 
the entire disk even with the planet tends to be less than $1$. After 
$15$ orbits, the optically thin temperature profiles range from 
adiabatic to isothermal depending on the value of $\beta$. 
\begin{figure}[h]
\begin{center}
\plotone{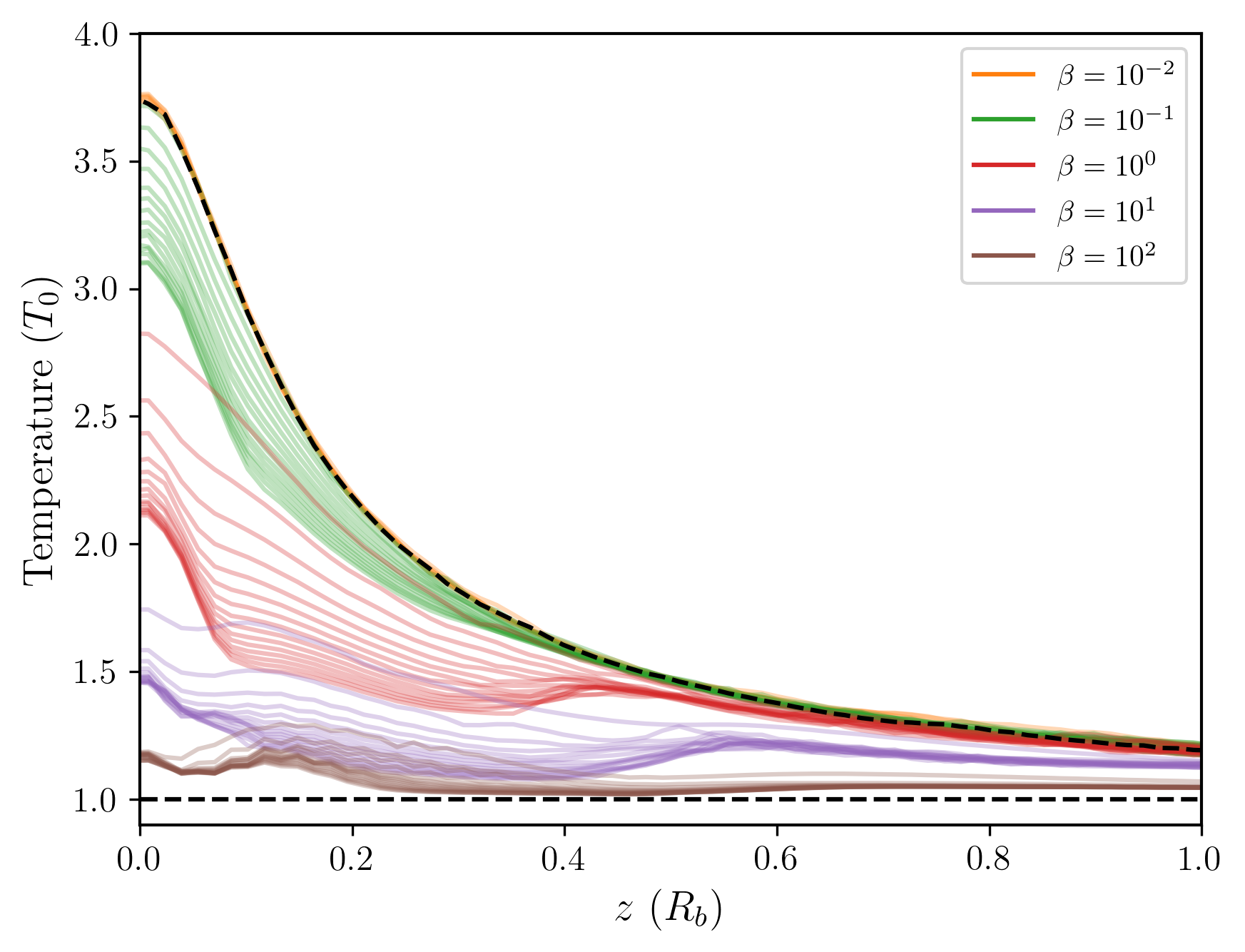}
\caption{Evolution of vertical temperature profiles along the low-opacity branch. Lines plotted roughly every orbit for 15 orbits. Temperature profiles decrease as atmospheres radiatively cool}\label{fig:temp01}
\end{center}
\end{figure}
These temperature profiles are plotted in Fig.\@ \ref{fig:temp01} 
in the manner of Fig.\@ \ref{fig:jupvert}. The temperature profiles 
are once again bounded by the adiabatic and isothermal models, 
becoming more adiabatic for low $\beta$ and more isothermal for 
large $\beta$. This ordering can be understood in terms of the 
optically thin cooling time in a particular model. A thermal 
perturbation about a homogenous equilibrium solution in an optically 
thin medium decays at a characteristic rate 
\citep{UnnoSpeigel1966}:
\begin{equation}
    t_{\rm cool}^{-1} = \frac{c_\gamma}{\lambda} \sim 
    \kappa\beta\Omega_0
\end{equation}
This suggests that in our optically thin simulations, the ratio of 
dynamical time to cooling time is simply the product of dimensionless 
parameters $\kappa\beta$. When the dynamical time and cooling time are 
roughly equal $(\beta=10^{2})$, the envelope is subject to rapid 
cooling over the course of our $15$ orbit simulation and the result 
is a nearly isothermal temperature profile. For lower values of 
$\beta$, the cooling time becomes longer than the dynamical time 
and the temperature profile appears increasingly adiabatic. 

\subsubsection{Mass accretion}
\begin{figure}
\begin{center}
\plotone{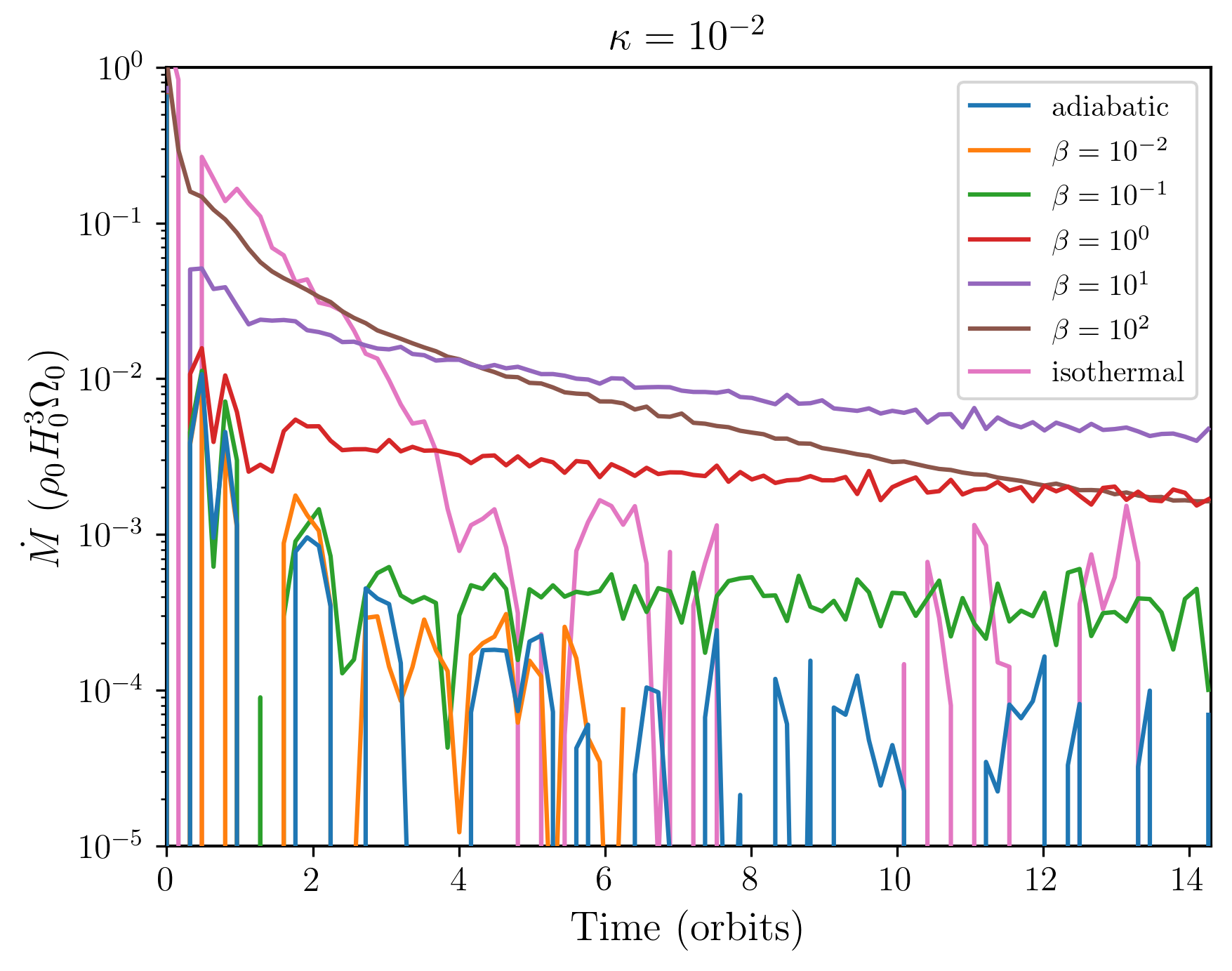}
\caption{$\dot{M}$ for mass inside Bondi radius as a function of time for models along the optically thin branch.}\label{fig:m01}
\end{center}
\end{figure}

The ordering seen in the temperature profiles also manifests in 
the magnitude of accretion rates. In 
Fig.\@ \ref{fig:m01}, we plot the rate of change in mass of the Bondi 
radius over the simulations' $15$ orbits.
Whereas the isothermal and adiabatic 
models reach steady state with accretion rates hovering around zero, 
optically thin radiative models tend to continually cool and accrete
for the full simulation lifetime. The rate at which they accrete 
mass also appears to be determined by the optically thin cooling rate 
$\kappa\beta$. Increasing $\beta$ by an order of magnitude, roughly 
increases the accretion rate by the same amount. Exceptions to this 
occur for the lowest value of $\beta=10^{-2}$ and the highest 
value of $\beta=10^{2}$. The former case has a fairly noisy accretion 
rate close to zero, similar to the adiabatic case. The accretion 
rate in this case may be too low to be distinguishable from noise 
at this resolution. For $\beta=10^2$, the radiative timescale and 
dynamical timescale are comparable making the envelope nearly 
isothermal. Departures from isothermality still appear great enough 
to slow the evolution to a still-accreting state by the end 
of $15$ orbits however. 

For these optically thin models, we also find that it is not entirely 
necessary to solve the full transfer equation. By running a set 
of models with a simple Newtonian cooling prescription, we find 
good agreement with our fully radiative models when the Newtonian 
relaxation time was set to the product $\kappa\beta$.

\subsection{Optically Thick Models}
\subsubsection{Envelope Structure}
\begin{figure}[h]
\begin{center}
\plotone{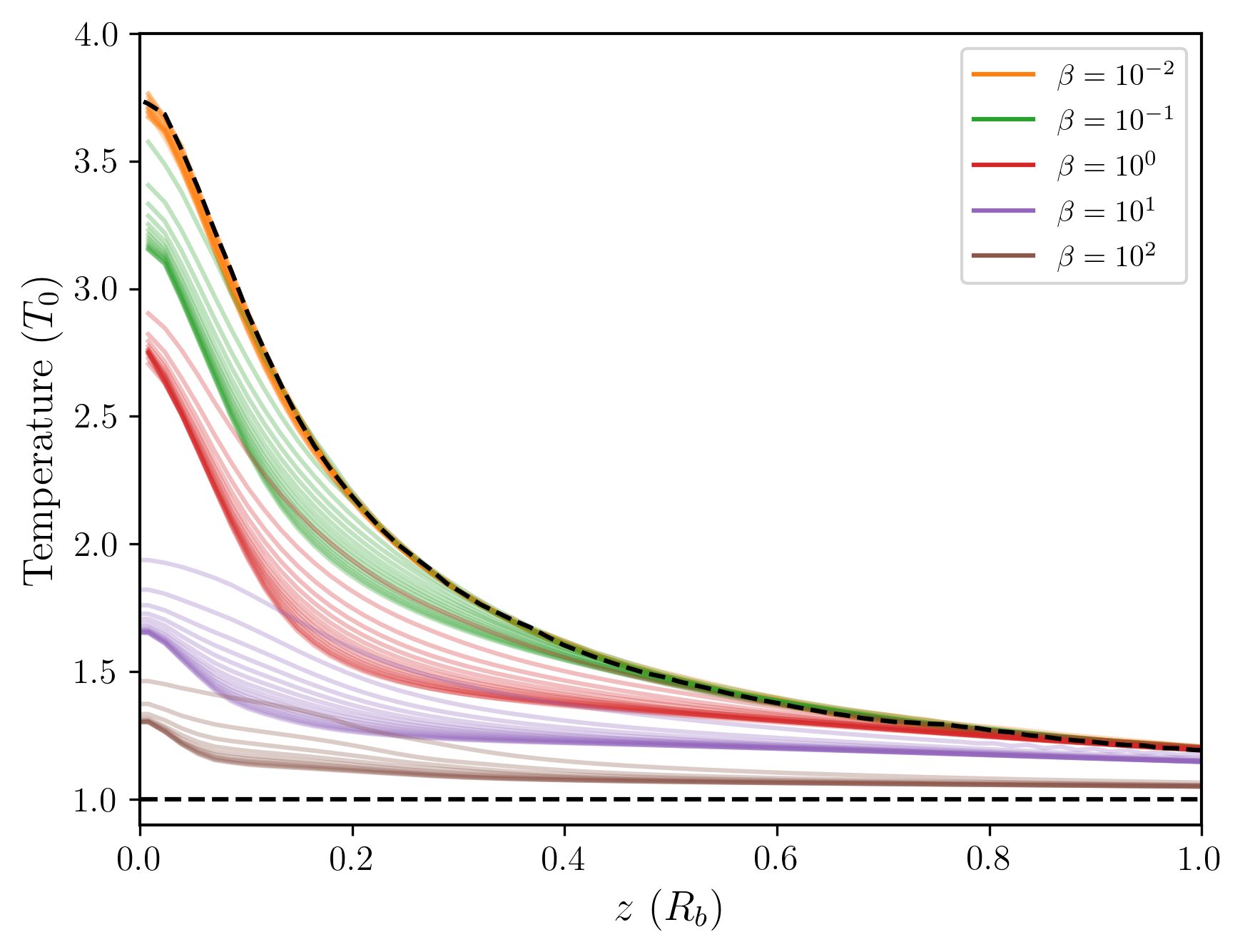}
\caption{Evolution of vertical temperature profiles along the optically thick branch. Lines plotted roughly every orbit for however long simulation ran. Temperature profiles decrease as atmospheres radiatively cool}\label{fig:temp100}
\end{center}
\end{figure}

These models, occupying the rightmost branch of Fig.\@ \ref{fig:par} 
are challenging to run for long as the explicit method timestep 
becomes prohibitive, particularly as one goes to low $\beta$. 
For this reason, we employ the implicit method for runs with 
$\beta \leq 1$. As mentioned in the examination of our 
fiducial proto-Jupiter model, an explicit run is also performed for 
$\beta=1$ to confirm that the methods agree in overlapping regions 
of parameter space. We plot the vertical temperature profiles for this 
branch in Fig.\@ \ref{fig:temp100}. 
Similar to the optical thin branch, optically 
thick models span a range from nearly isothermal to nearly adiabatic 
depending on their value of $\beta$. More than that, for a given 
value of $\beta$, models on the optically thin and optically thick 
branch relax to similar temperature profiles. The similarity 
can be understood by considering the optically thick cooling timescale. 
For a perturbation of wavenumber $k$ on a homogenous optically thick 
medium, the cooling rate is \citep{UnnoSpeigel1966}:
\begin{equation}
    t_{\rm cool}^{-1} = c_\gamma k^2 \lambda \sim \frac{\beta}{\kappa}\frac{k^2}{\Omega_0}
\end{equation}
Because the optically thick cooling rate scales as $\beta/\kappa$, 
a model on our optically thick branch with given $\beta$ has roughly 
the same dimensionless cooling time as a model on the optically 
thin branch with the same $\beta$. 

\subsubsection{Mass accretion}
\begin{figure}[h]
\begin{center}
\plotone{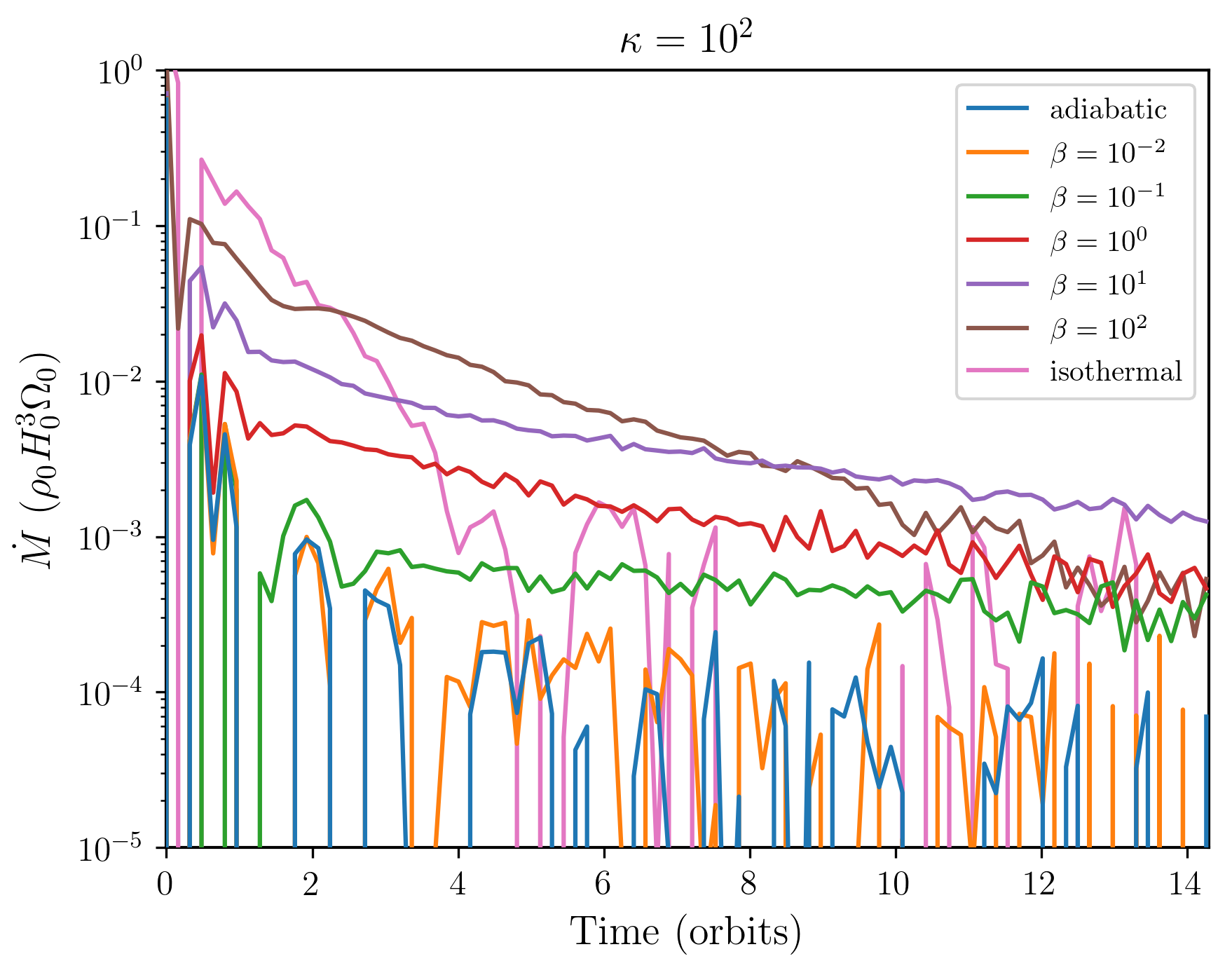}
\caption{$\dot{M}$ for mass inside Bondi radius as a function of time for various models along the optically thick branch.}\label{fig:m100}
\end{center}
\end{figure}

Mass accretion rates for optically thick models (Fig.\@ 
\ref{fig:m100}) 
are roughly similar to their optically thin 
counterparts, at least in terms of their ordering. 
However, optically 
thick models with $\beta\geq 1$ show more temporal variation 
than their optically thin counterparts, decaying by roughly an 
order of magnitude over the $15$ orbits. The $\beta < 1$ models 
are nearly identical between the $\kappa=10^{-2}$ and $\kappa=10^2$, 
showing low time-independent accretion rates. 

\subsection{$\beta=1$ Models}
\subsubsection{Envelope Structure}
\begin{figure}[h]
\begin{center}
\plotone{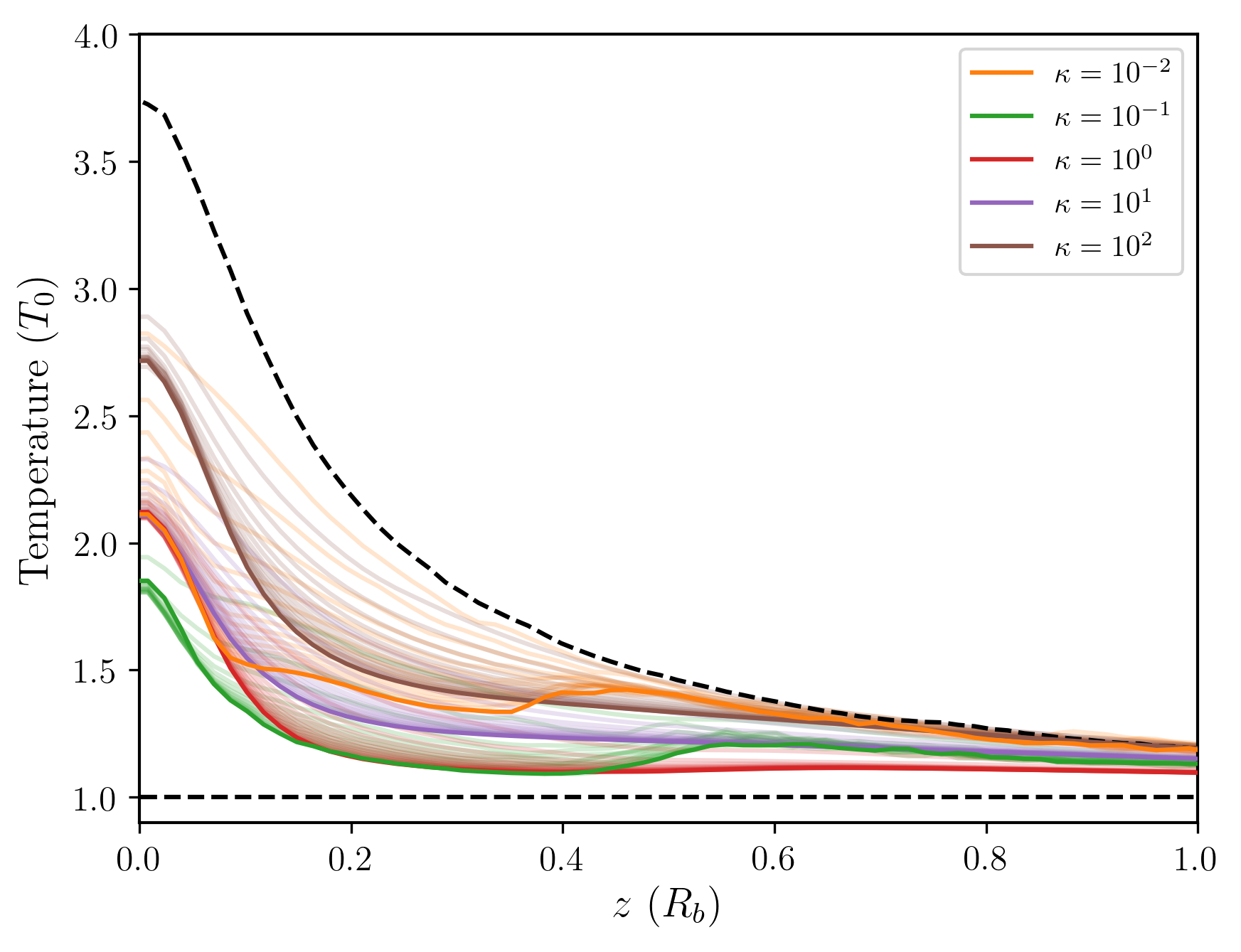}
\caption{The vertical temperature profiles for models at fixed $\beta=1$. Unlike Figs.\@ \ref{fig:temp01} \& \ref{fig:temp100}, final profiles are plotted as lines with increased visual opacity to help distinguish between overlapping models.}\label{fig:beta1temp}
\end{center}
\end{figure}

These models are performed at fixed $\beta$, but variable opacity. 
This is more consistent with other studies where the strategy is 
to usually adopt some fiducial background conditions and then 
vary the model opacity. In a simple sense, this allows models 
to parameterize uncertainties associated with dust properties and 
settling mechanisms in the protoplanetary disk. Unlike the models 
at fixed $\kappa$, these models with identical $\beta$ share similar 
vertical temperature profiles which we plot in Fig.\@ 
\ref{fig:beta1temp}. 
The temperature profiles are not ordered 
monotonically as they are in optically thick \& thin branches. 
While the most optically thin and optically thick models tend to be 
slightly hotter, the $\beta=10^{-2}$ model also matches the onto the 
temperature profiles of the $\beta=10^{-1}$ and $\beta=10^1$ models 
at small radii. This may be in part due to the softening length 
as the overlap emerges around $0.1 R_b$.

\subsubsection{Mass Accretion}
\begin{figure}[h]
\begin{center}
\plotone{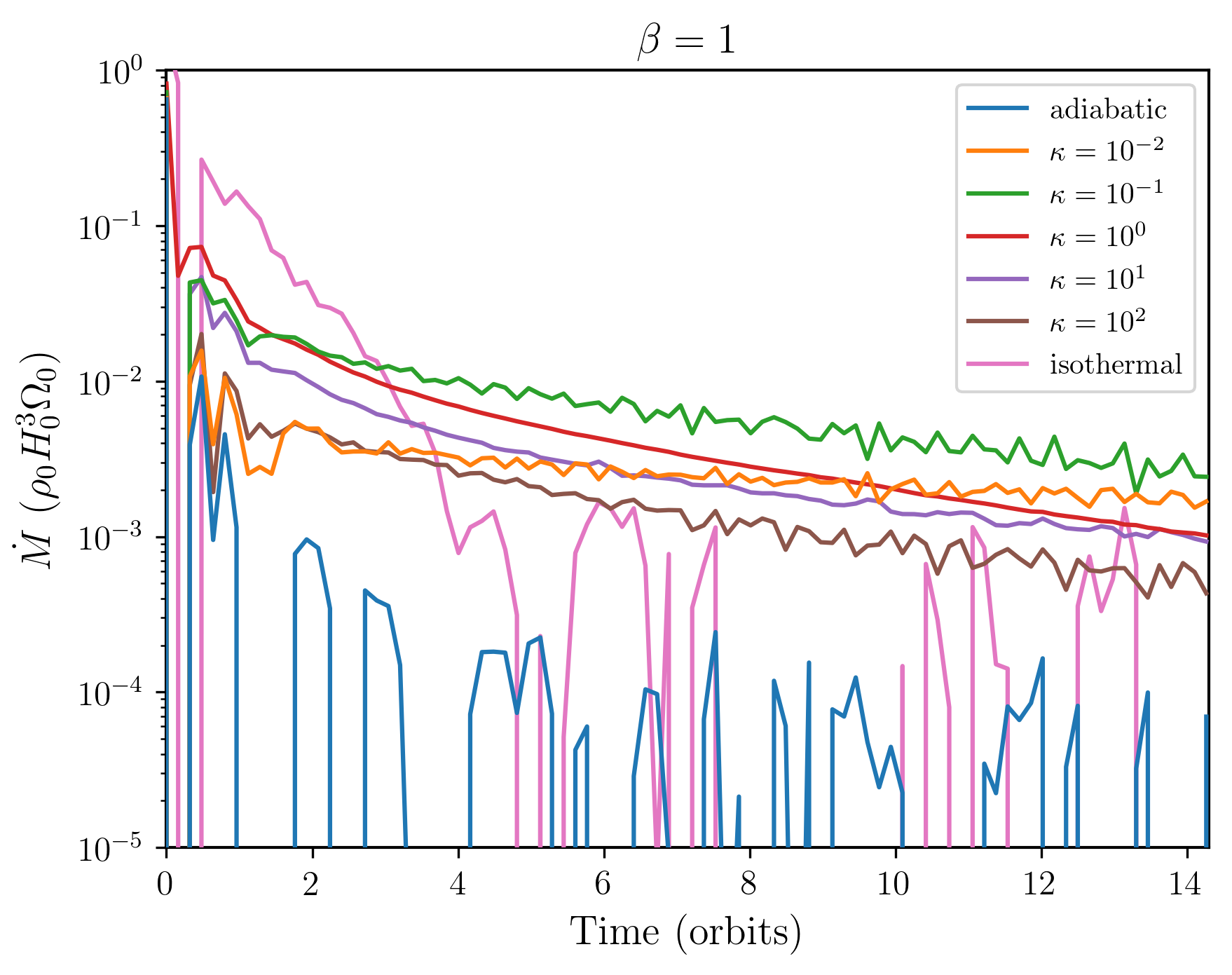}
\caption{Mass accretion rate at fixed $\beta=1$ but varying opacity.}\label{fig:beta1}
\end{center}
\end{figure}

Similar to the temperature profiles, mass accretion rates are 
more clustered than the optically thin and thick branches, 
suggesting a fairly weak scaling of $\dot{M}$ with $\kappa$. 
As before, we display the accretion rates in the Bondi sphere in 
Fig.\@ \ref{fig:beta1}.
While models with similar cooling times (e.g. $\kappa=10^{-2}$ and 
$\kappa=10^{2}$) begin accreting at equal rates, once the planet 
has accrued something of an envelope, the optically thin accretion 
rates decay more slowly. The result is a higher accretion rate 
for optically thin models after $15$ orbits. While the 
mass accretion rate does not vary monotonically with opacity because 
the $\kappa=10^{-1}$ model has higher accretion rate than 
$\kappa=10^{-2}$, the $\kappa=10^{-1}$ accretion rate is decaying 
more quickly. If this trend continues, then a monotonic variation 
of accretion rate with opacity is likely given longer simulation 
runtimes.

\section{Discussion}
\subsection{Adiabatic or Isothermal?}
Depending on the value of $\beta$, the models span the space from 
isothermal to adiabatic temperature profiles by formal end of our 
simulations. The precise state of gas in the outer envelope is 
important because the planetary flow field is linked to the 
assumed gas thermodynamics. Both the character of recycling flows 
\citep{Kurokawa+2018} and CPD formation \citep{Fung+2019}
can vary greatly depending on the assumed equation of state. 
The opacity $\kappa$ does have some minor effect on the 
outer envelope temperature profiles, with models generally becoming 
more adiabatic towards higher optical depth. However, the envelope 
temperature profiles appear to be primarily determined by the choice 
of model $\beta$. While it is colloquially thought that optically 
thin atmospheres should be isothermal and optically thick 
atmospheres adiabatic, we demonstrate the existence of optically 
thick atmospheres with nearly isothermal temperature profiles and 
rapid cooling. This suggests isothermal models, 
often discounted for being highly idealized and unrealistic, may 
occupy some reasonable region of parameter space. Because isothermal 
models contain robust CPDs, this further suggests that CPDs could 
exist around low-mass planets, given large enough $\beta$. 
The models presented here adopt a large softening length, resulting 
in artificially low rotational velocities and 
making it difficult to investigate this question of CPD formation. 
We have run preliminary models with smaller softening length but 
have encountered numerical issues and departures from symmetry for 
models along the optically thick branch. Models running along the 
optically thin branch do suggest the formation of CPDs for 
$\beta \gtrsim 10$ and disappearance for the more adiabatic looking 
$\beta \lesssim 1$.

It is standard to take models after several tens of orbits 
and treat them as representative of the envelope state. 
Nevertheless, there remains some question of how well this translates 
to actual planets as the formation process is orders of 
magnitude longer. 
1D static models, having assumed a diffusion 
approximation everywhere, typically find a more isothermal profile 
in the outer radiative envelope than the optically thick models 
run here. This suggests that our models may need to be run longer 
to truly compare to 1D models. 
Were we to take these models and continue running them as they are 
however, we would expect them to simply continue cooling to an 
isothermal profile ad infinitum. 
While this is true of something like Jupiter that is still cooling 
and contracting to this day, it is unrealistic in the sense
that the gravity of our planet does not increase with the 
addition of mass. In the true planet-forming case, the accretion 
rate would not continue to decrease indefinitely because 
more massive envelopes accrete at higher rates. Eventually 
the accretion rate would be so large as to be termed runaway, 
but our models, being fixed potentials, could never experience 
a runaway phase. 
We have also provided no sources of heating either due to 
planetesimal accretion or an interior envelope in these models. 
While these would only act to slow the accretion process, their 
absence also leads to the indefinite cooling of our models.
In the future, we hope to 
improve our models by imposing a luminosity near the softening 
length to represent the subgrid interior envelope. Depending on 
the cooling time, the heated outer envelope would presumably 
come to a steady state and could be compared more directly with 
static 1D models. 

\subsection{Accretion Rates}
While we have presented model accretion rates in the preceding 
sections, we do not regard these as actual planetary accretion rates.
For one, these are the accretion rates of mass within the planet's 
Bondi radius and there is no guarantee that all of this mass remains 
bound to the planet. Furthermore, the mass accretion rates have 
a tendency to decrease with time, suggesting that accretion rates 
could be made lower by simply adopting a longer runtime. For these 
reasons we do not treat the presented rates as indisputable but 
rather as upper bounds on some true planetary accretion rate so long 
as the simulations are converged.  
Though our models also adopt a larger softening length than is 
realistic for most planets, the results of \citet{Schulik+2019} 
suggest that going to smaller softening length only decreases 
the corresponding accretion rates. In particular, lowering 
the softening length creates a deeper potential thereby raising 
the energy content of the envelope. While the luminosities 
also increase, it is not in keeping with the energy content, 
resulting in a lower net mass accretion rate. This suggests that 
the interpretation of our model accretion rates as an upper bound 
is robust. 

\begin{figure}[h]
\begin{center}
\plotone{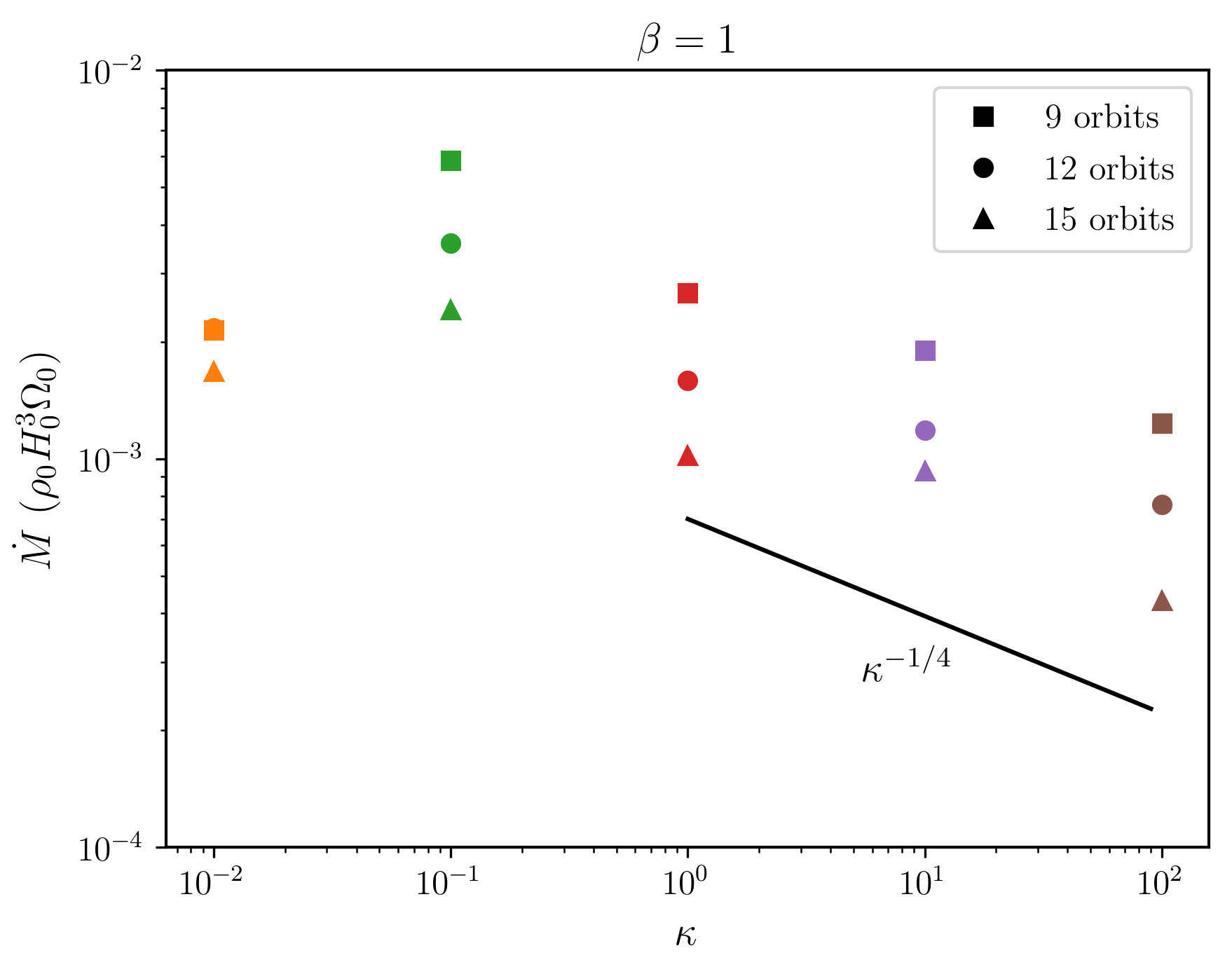}
\caption{Accretion rates as a function of model opacity. Points are sampled three times from the simulation at $9$ orbits (squares), $12$ orbits (circles) and $15$ orbits (triangles). We find that accretion scale as roughly $\kappa^{-1/4}$, with $\kappa=10^{-2}$ being a potential exception.}\label{fig:kapscale}
\end{center}
\end{figure}

Without being certain of the true planetary accretion rate, we 
are still in a position to make relative comparisons between 
the presented models and extend these to more physical scenarios.
In particular, we can infer from our $\beta=1$ branch some estimate of 
how the mass accretion rate varies with opacity. When we plot the 
$\dot{M}$ as a function of $\kappa$ for these models (Fig.\@ 
\ref{fig:kapscale}), we find an 
approximate $\kappa^{-1/4}$ scaling consistent with similar 
3D radiation-hydrodynamics studies that overlap our optically thick 
cases \citep{AyliffeBate2009, Schulik+2019}. We do find that our most 
optically thin model shows a departure from this scaling but 
due to the limited sampling in $\kappa$ it is difficult to 
tell how robust this is. Also because the mass accretion rate in the 
$\kappa=10^{-2}$ model exhibits less rapid decay than all the 
other models, it is also likely that points sampled from a longer 
simulation run would be more in accordance with the $\kappa^{-1/4}$ 
scaling. 
The $\kappa^{-1/4}$ scaling is distinct from 1D models 
that predict a scaling closer to $1/\kappa$ \citep{Ikoma+2000}. 
Other studies investigating higher mass planets have found 
the $\kappa$ dependence continues to weaken as the accretion 
becomes disk-limited \citep{DangeloBodenheimer2013}. While it 
has been suggested that $\kappa^{-1/4}$ is simply an intermediate 
regime lying between the 1D and disk-limited rates, our models here 
are low enough mass ($10 M_\oplus$) as to overlap with the 1D regime. 
This suggests a more fundamental difference between the 1D and 
3D models that we hope to investigate in the future by directly 
comparing with 1D models. Other studies with this sub-linear 
opacity scaling have suggested that the dependence could be 
reflective of model initial conditions as their methodology changes 
the initial disk midplane density and opacity simultaneously 
\citep{Schulik+2019}. Because we use a local box with the same 
initial conditions independent of opacity, our simulations suggest 
that this scaling is more robust and we are at least able to rule 
out that hypothesis in the meantime.
\begin{figure}
\plotone{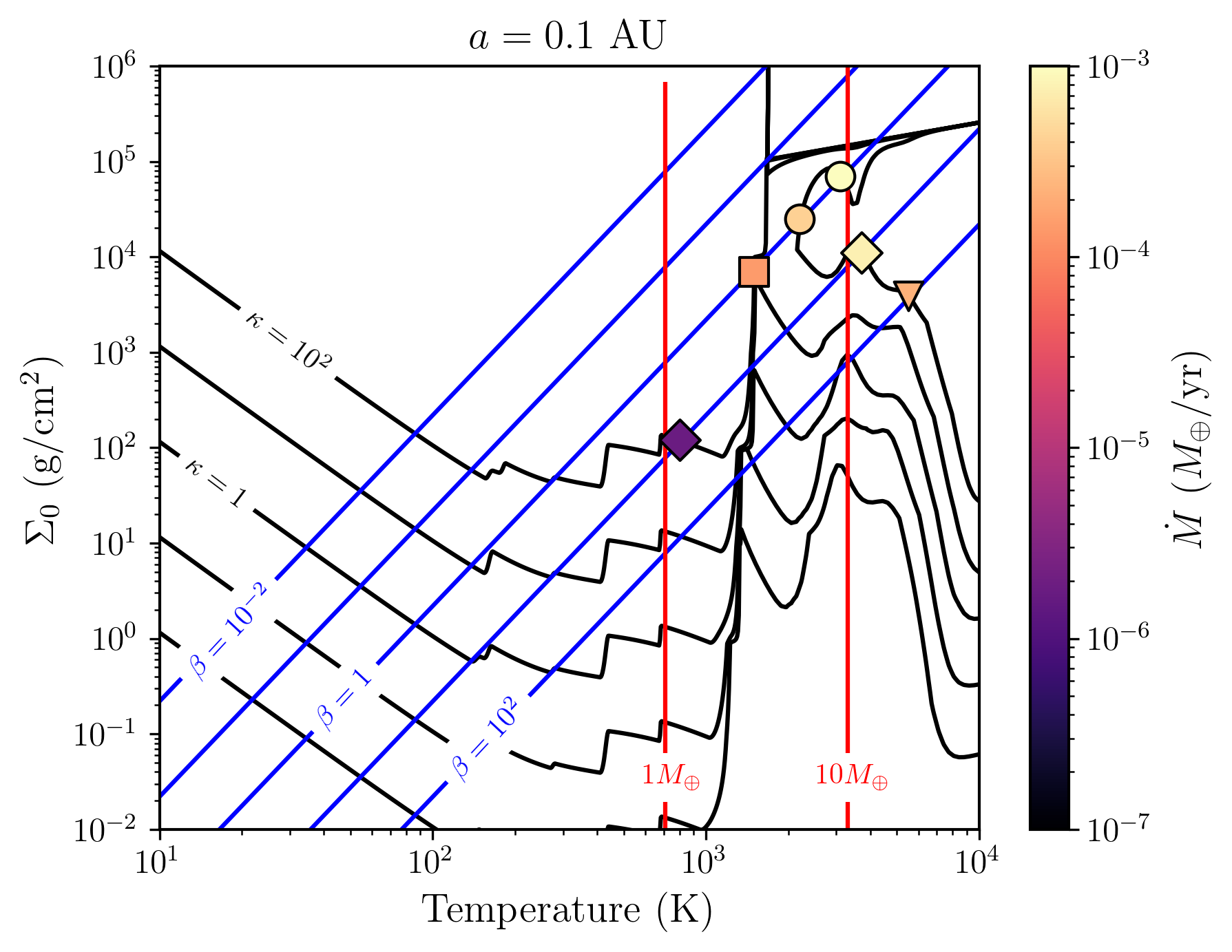}
\caption{Dimensional mass accretion rates at $0.1$ AU as a function 
of disk surface density and disk temperature. Black lines are contours 
of $\kappa = \left[10^{-2},10^{-1},10^{0},10^{1},10^{2}\right]$ 
assuming \citet{Semenov+2003} opacities for 
$\kappa_0\left(\rho_0,T_0\right)$. Blue lines are contours of 
$\beta = \left[10^{-2},10^{-1},10^{0},10^{1},10^{2}\right]$ The 
rightmost red line marks a $10 M_\oplus $ core and the left line 
a $1 M_\oplus$ core. At the contour intersections for particular 
values of $\kappa$ and $\beta$ we able to apply and dimensionalize 
our dimensionless model which we show with markers colored according 
to the model accretion rate.
}
\label{fig:0.1au}
\end{figure}

\begin{figure}
\plotone{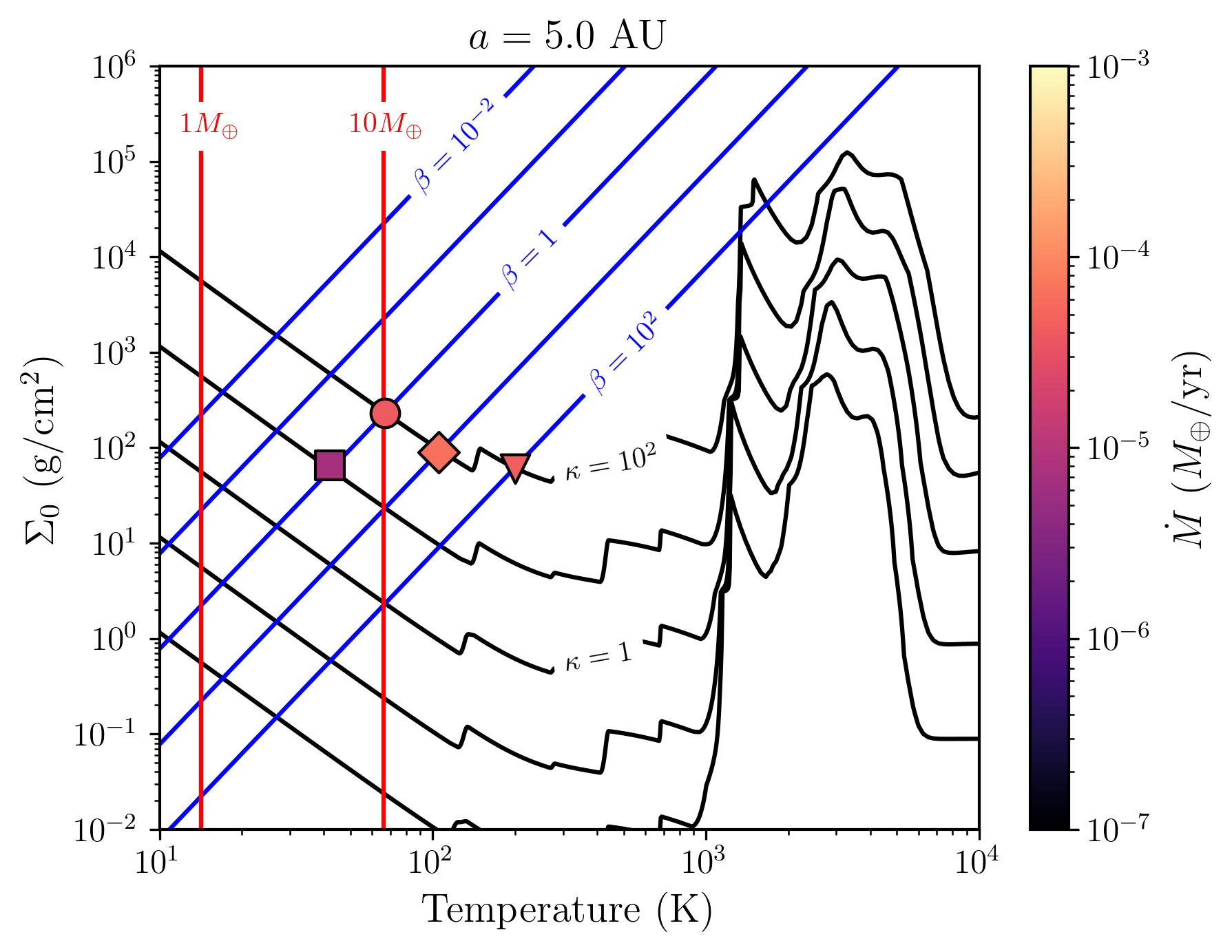}
\caption{Dimensional mass accretion rates at $5.0$ AU as a function 
of disk surface density and disk temperature plotted in the 
style of Fig.\@ \ref{fig:0.1au}. Models sharing the same 
dimensionless $(\beta,\kappa)$ are plotted with the same symbol 
between this figure and Fig.\@ \ref{fig:0.1au}.
}
\label{fig:5.0au}
\end{figure}
For some set of conditions in the disk, 
like $(\rho_0,T_0,\kappa_0,a)$, 
it is possible to dimensionalize our models and study 
mass accretion throughout the disk. Because we choose 
a fixed thermal mass $q_t=0.5$, the planet mass of a given model 
is fixed by choosing a disk temperature $T_0$ and 
an orbital radius $a$. While it is unfortunate that we may not 
vary a planet's mass and disk conditions independently, we have 
chosen $q_t$ in particular to be applicable to both proto-Jupiter 
and super-Earth conditions. Dimensionalizing also requires some 
choice of the opacity $\kappa_0$ which we take to be the 
temperature and density dependent opacities of \citet{Semenov+2003}.
While this is not entirely self-consistent as the models themselves 
were simulated with a temperature and density independent opacity, 
it is more informative than arbitrarily choosing some constant 
opacity. With $\kappa_0(\rho_0,T_0)$, we can choose a location in 
the disk $a$, a density $\rho_0$, and a temperature $T_0$ and 
calculate the dimensionless $(\beta,\kappa)$. In Figs.\@ 
\ref{fig:0.1au} \& \ref{fig:5.0au}, we invert this process, 
taking our models which have a fixed $(\kappa,\beta)$ and finding 
the $(\rho_0,T_0)$ that these models could represent in a physical 
disk. We then dimensionalize the mass accretion rate and plot 
them as a function of $(\Sigma_0,T_0)$. Fig.\@ \ref{fig:0.1au} 
corresponds to the choice of $a=0.1$ AU and as the points lie 
roughly in the $1-10M_\oplus$, $T_0\sim 1000$ K, 
$\Sigma_0\sim 10^5$ g/cm$^2$ region, they are indeed super-Earth 
like. Fig.\@ \ref{fig:5.0au} 
corresponds to the choice of $a=5.0$ AU and models are applicable to 
Jupiter-like conditions at $T_0\sim100$ K, $\Sigma_0\sim 100$ 
g/cm$^2$.

With the dimensional accretion rates in Figs.\@ \ref{fig:0.1au} \& 
\ref{fig:5.0au}, we notice that the magnitude of accretion rates 
is large enough such that all models would be able to accrete an 
envelope comparable to their core mass within a disk lifetime 
$(t\sim 10^6)$ and potentially enter a runaway stage of growth. 
However this quoted accretion rate is only representative of the 
earliest phases of planet growth. On scales longer than the 
simulated time here and for most of the protoplanet's lifetime, 
we expect the accretion rate to be significantly lower. 
In order to estimate the true accretion rate 
for the pre-runaway lifetime it becomes necessary to construct  
a 1D model consistent with and informed by the 3D simulation 
in the regime where they overlap and then evolve the 1D on longer 
timescales. Rather than do this here, we refer to an additional paper 
\citep{BaileyZhu2023} focused on adequately parameterizing the 
effects of 3D recycling into 1D models. Here instead we focus on the 
model setup, fidelity, details, and dimensionless framework.

\subsection{Validity of Radiative Transfer Approximations}
In previous sections, we remarked on and showed some evidence of the 
agreement between the implicit/explicit methods. Because the 
explicit method neglects the time derivative of $I$ and any 
velocity dependent terms in the solution of the radiation field, 
it is a static approximation. In particular, fluid motion must occur 
more slowly than the time it takes radiation to diffuse or free 
stream. The implicit method on the other hand solves the full 
velocity-dependent transfer. Due to the good agreement between 
the two methods we can conclude that static methods are perfectly 
viable and computation of velocity dependent terms is extraneous 
for these types of models.

\begin{figure*}
\plotone{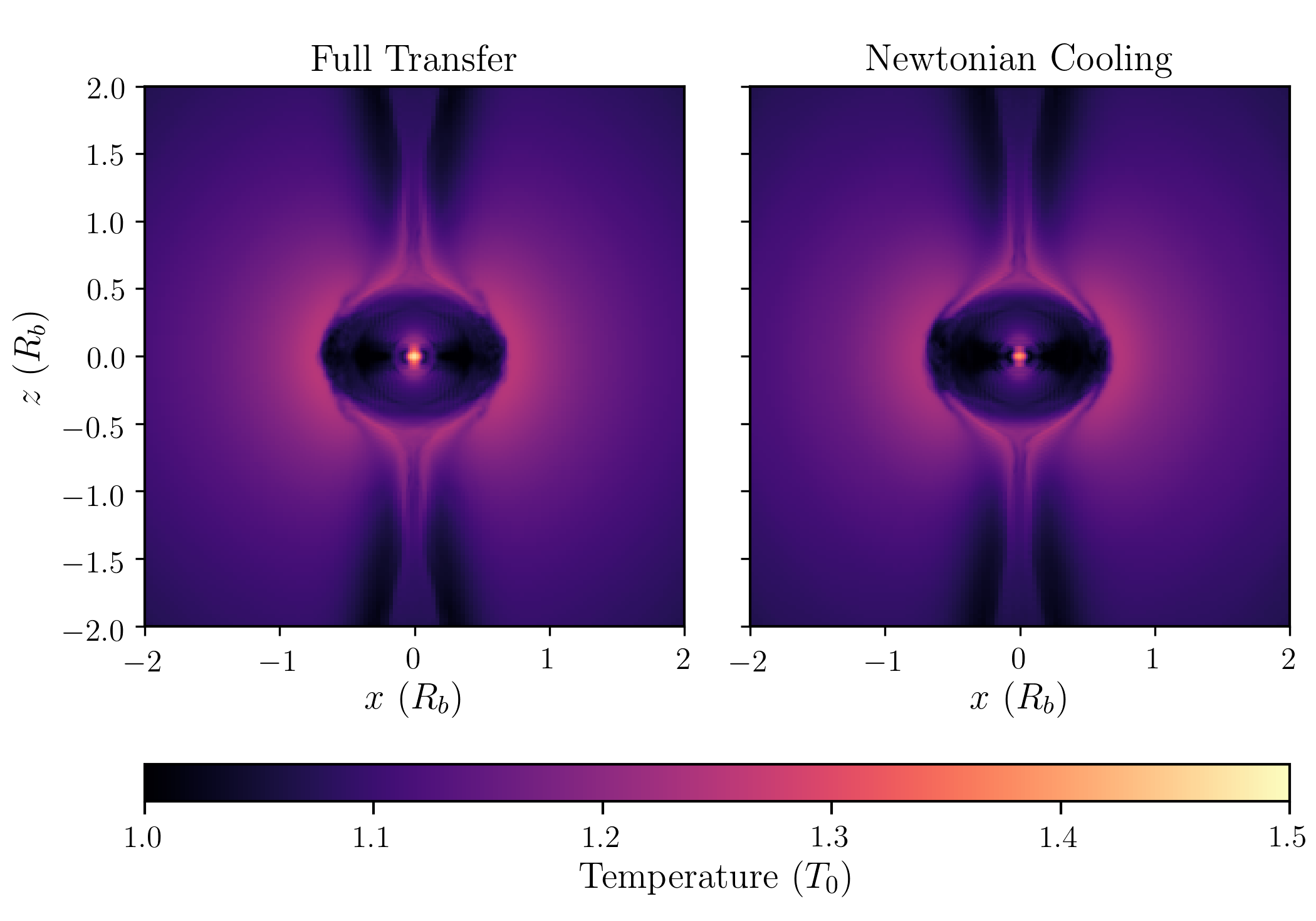}
\caption{Vertically sliced temperature profiles after $15$ orbits 
for the optically thin $\beta=10$ model. The left panel comes from a 
simulation employing our full radiative transfer module whereas the 
right panel simply applies a linear cooling law as a source term 
to the energy equation. 
}
\label{fig:newton}
\end{figure*}

Because we utilize methods that solve the transfer equation instead 
of a closure approximation, we are also in a position to evaluate the 
fidelity of radiative approximations commonly employed by other 
studies. The simplest cooling model is a Newtonian linear cooling law, 
applicable to optically thin regions. For our optically thin 
$\kappa=10^{-2}$ models, we find remarkable agreement with a 
simple linear cooling law. Fig.\@ \ref{fig:newton} shows a 
comparison of the sliced temperature profiles of two 
optically thin models, one solving the full transfer equation 
and one applying linear cooling as an energy source term. 
So long as the linear cooling rate is set appropriately 
($t_{\rm cool}\sim \kappa\beta$), we find similar agreement for all 
models on our optically thin branch. While this is convenient because 
a linear cooling law adds negligible computational cost, the 
applicability to planet simulations is rather slim. As suggested 
by Fig.\@ \ref{fig:par}, only at large orbital radius (i.e. low 
surface density) is optically thin reasonable and even then 
the density increase near the planet can make some part of 
the envelope optically thick. For that reason, this method is 
only really tenable when focused on an optically thin outer 
envelope or only interested in a toy model for envelope cooling e.g. 
\citet{Kurokawa+2018}.

Though we can't explicitly perform FLD runs as it is not 
currently implemented in \textsc{athena++}, we are able to 
examine the true fluxes and determine if they are consistent with 
a diffusion approximation.
\begin{figure*}
\plotone{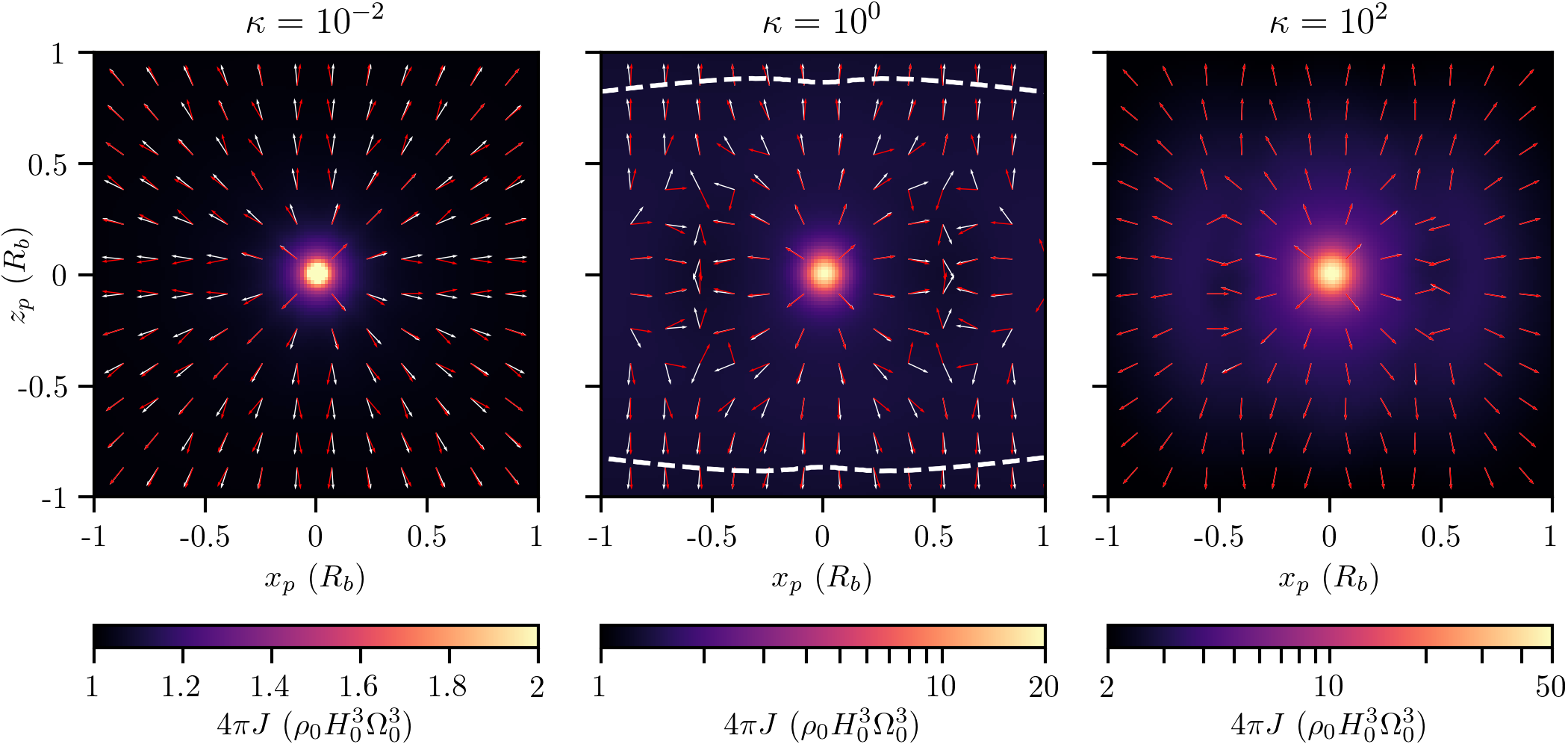}
\caption{Vertically sliced profiles of the radiation field for 
different opacity models. Slices are colored 
according to the mean intensity $J$. White vectors show the true 
radiation flux while red vectors show the gradient of the radiation 
energy density. All are unit vectors and not representative of the 
magnitude. The dashed line marks the location of vertical 
optical depth $\tau=1$. While $\kappa=10^{-2}$ is too optically thin 
to reach unity optical depth, the $\tau=1$ surface for $\kappa=10^2$ 
lies outside the range of the plot at a height of nearly $4R_b$.
}
\label{fig:er_zoom}
\end{figure*}
In Fig.\@ \ref{fig:er_zoom}, we plot 
the mean intensity (radiation energy density up to a constant) 
for several different opacity models on the $\beta=1$ branch. 
Direction vectors for the radiation flux and also the gradient 
of the energy density $\nabla E_r$ are also shown and match very 
well for the optically thick model. For the optically thin model, 
while vectors don't match as perfectly, both the flux and the energy 
density gradient are nearly radial. In the intermediate $\kappa=1$ 
case, the fluxes are no longer radial, showing interesting deviations 
based on temperature variations in the planetary envelope. While 
the fluxes and $\nabla E_r$ can be largely misaligned in certain 
areas of the envelope, the vectors end up being roughly aligned 
and largely vertical on the scale of the photosphere. This 
suggests that FLD, particularly in the optically thick models 
should be fairly reasonable. Considering that solution of 
the transfer equation tends to add 1-2 orders of magnitude in 
computation time depending on the number of angles per cell, we 
recommend the use of FLD for this application. In particular, 
we believe the advantages offered by going to higher resolution 
and longer runtimes far outweight the benefits offered by more 
accurate radiative transfer.

\section{Summary}
We have carried out a range of hydrodynamics 
simulations of envelopes around planetary cores employing a 
solution of the full radiative transfer equation. We formulate 
the problem in terms of dimensionless parameters $\kappa$, $\beta$, 
$q_t$, and $\epsilon$ (Sec.\@ \ref{sec:radsim}). Simulations 
were performed for a variety of $\kappa$ and $\beta$ corresponding 
to different disk conditions and optical depths ranging from 
optically thin to optically thick. With these simulations we were 
able to obtain profile for envelope structure and mass accretion 
rates or at least upper limits on the minimum accretion rate. We list 
some of our findings:
\begin{itemize}
    \item Planet envelopes with radiative cooling are neither 
    adiabatic nor isothermal. Precisely how adiabatic or isothermal 
    they are is not solely a function of opacity but is primarily 
    determined by $\beta$, related to the cooling time. 
    \item Models show a weak $\kappa^{-1/4}$ scaling of accretion rate well below the disk-limited accretion regime. This is less 
    steep than reported by 1D evolutionary models at similar stages. 
    \item We have obtained upper-limits to 3D mass accretion rates 
    prior to runaway growth for a wide range of parameters. This corresponds to lower-limits on 
    the formation timescale or time to reach crossover mass. 
    The rates are dimensionalized for proto-Jupiters and super-Earth 
    like formation conditions in Figs.\@ \ref{fig:0.1au} \& 
    \ref{fig:5.0au}.
    \item Flux limited diffusion is a reasonable method for simulating 
    planetary envelopes, particularly when one considers the 
    cost savings over more sophisticated methods. 
\end{itemize}

\begin{acknowledgements}
The authors are also pleased to acknowledge that the work reported on in this paper was substantially performed using the Princeton Research Computing resources at Princeton University which is consortium of groups led by the Princeton Institute for Computational Science and Engineering (PICSciE) and Office of Information Technology's Research Computing.

The simulations presented in this article were performed on computational resources managed and supported by Princeton Research Computing, a consortium of groups including the Princeton Institute for Computational Science and Engineering (PICSciE) and the Office of Information Technology's High Performance Computing Center and Visualization Laboratory at Princeton University.
\end{acknowledgements}

\bibliography{planet_ecc}{}
\bibliographystyle{aasjournal}

\end{document}